\documentclass[twocolumn,prx,aps,amsmath,amssymb,longbibliography]{revtex4-1}

\usepackage{graphicx}
\usepackage{dcolumn}
\usepackage{bm}

\begin{document}

\title{Majorana finite frequency nonequilibrium quantum noise}

\author{Sergey Smirnov}
\affiliation{P. N. Lebedev Physical Institute of the Russian Academy of
  Sciences, 119991 Moscow, Russia}
\email{1) sergej.physik@gmail.com\\2)
  sergey.smirnov@physik.uni-regensburg.de\\3) ssmirnov@sci.lebedev.ru}

\date{\today}

\begin{abstract}
Quantum finite frequency noise is one of fundamental aspects in quantum
measurements performed during quantum information processing where currently
Majorana bound states offer an efficient way to implement fault-tolerant
quantum computation via topological protection from decoherence or unitary
errors. Thus a detailed exploration of Majorana finite frequency noise
spectra, preferably in a nonequilibrium device, is a timely challenge of
fundamental importance. Here we present results on finite frequency
differential noise that is the derivative of the noise with respect to the
frequency. This quantity has universal units of $e^2$ and scans in high detail
all peculiarities of the Majorana noise clearly demonstrating its universal
finite frequency features. In particular, we provide photon absorption spectra
on all energy scales and reveal a rich structure including universal Majorana
plateaus as well as universal Majorana resonances and antiresonances at
characteristic frequencies. Our results are of immediate interest to
state-of-the-art experiments involving quantum noise mesoscopic detectors able
to separately measure photon absorption and emission spectra.
\end{abstract}

\maketitle
\section{Introduction}\label{intro}
Majorana bound states in condensed matter setups \cite{Kitaev_2001} attract at
present a lot of attention from a fundamental point of view because they
partly mimic Majorana fermions \cite{Majorana_1937} in the elementary particle
physics. The topological superconducting phase of the Kitaev model
\cite{Kitaev_2001} in realistic setups
\cite{Alicea_2012,Flensberg_2012,Sato_2016,Lutchyn_2018} provides Majorana
bound states localized at the ends of a one-dimensional system. There is a
rather high degree of evidence that Majorana bound states have been observed
in transport experiments \cite{Mourik_2012,Albrecht_2016,Zhang_2018} measuring
the differential conductance. Nevertheless, fully conclusive experiments have
still to be performed. For example, thermodynamic experiments able to detect
Majorana tunneling entropy \cite{Smirnov_2015} do not require any braiding
operation and at the same time are fully conclusive with respect to the
presence of Majorana bound states. Recent experimental progress in direct
entropy measurements \cite{Hartman_2018} within a mesoscopic setup clearly
demonstrates that such thermodynamic detections of Majorana bound states are
quite feasible and might appear soon.

Practically, non-Abelian Majorana zero-energy modes provide a platform for
topological quantum computation \cite{Kitaev_2003} as a sequence of unitary
transformations implementing braid operations. Since these unitary evolutions
depend only on the topological properties of the resulting braids, they are
immune to local perturbations. This protects, {\it e.g.}, quantum storage
devices from decoherence errors induced by certain interactions with an
environment \cite{Nayak_2008}. Possible experiments on braiding Majorana zero
modes have been theoretically proposed in setups with $T$-junctions
\cite{Alicea_2011,Aasen_2016}. Braiding schemes avoiding unitary evolutions
and using instead a projective measurement are also available via, {\it e.g.},
an electron teleportation \cite{Vijay_2016}.

Noise plays a twofold role in Majorana topological computations. First,
although braids are topologically protected, qubits may loose their coherence
when Majorana bound states move during a braiding operation in the presence of
noise induced by an environment \cite{Pedrocchi_2015}. Second, qubit readout
schemes often involve electric currents \cite{Vijay_2016,Gharavi_2016} whose
noise governed by Majorana degrees of freedom is in general finite at all
frequencies. This finite frequency noise is able to encode the Majorana qubit
state for a readout by a proper quantum noise detector able to reach the
quantum limit \cite{Clerk_2010}. Moreover, noise itself may be viewed as a
unique Majorana signature extracted from advanced transport experiments
alternative to the differential conductance.

Majority of research on Majorana noise in mesoscopic setups is focused on
symmetrized noise \cite{Liu_2015,Liu_2015a,Beenakker_2015,Haim_2015} which is
often explored in the static limit, {\it i.e.} at zero frequency. A
characteristic feature of zero frequency nonequilibrium Majorana noise is its
universal behavior leading, {\it e.g.}, to universal effective charges
\cite{Smirnov_2017} equal to $e/2$ and $3e/2$ at low and high bias voltages,
respectively, or to universal ratios between nonlinear response coefficients
of Majorana thermoelectric noise and mean current \cite{Smirnov_2018}.

\begin{figure}
\includegraphics[width=8.0 cm]{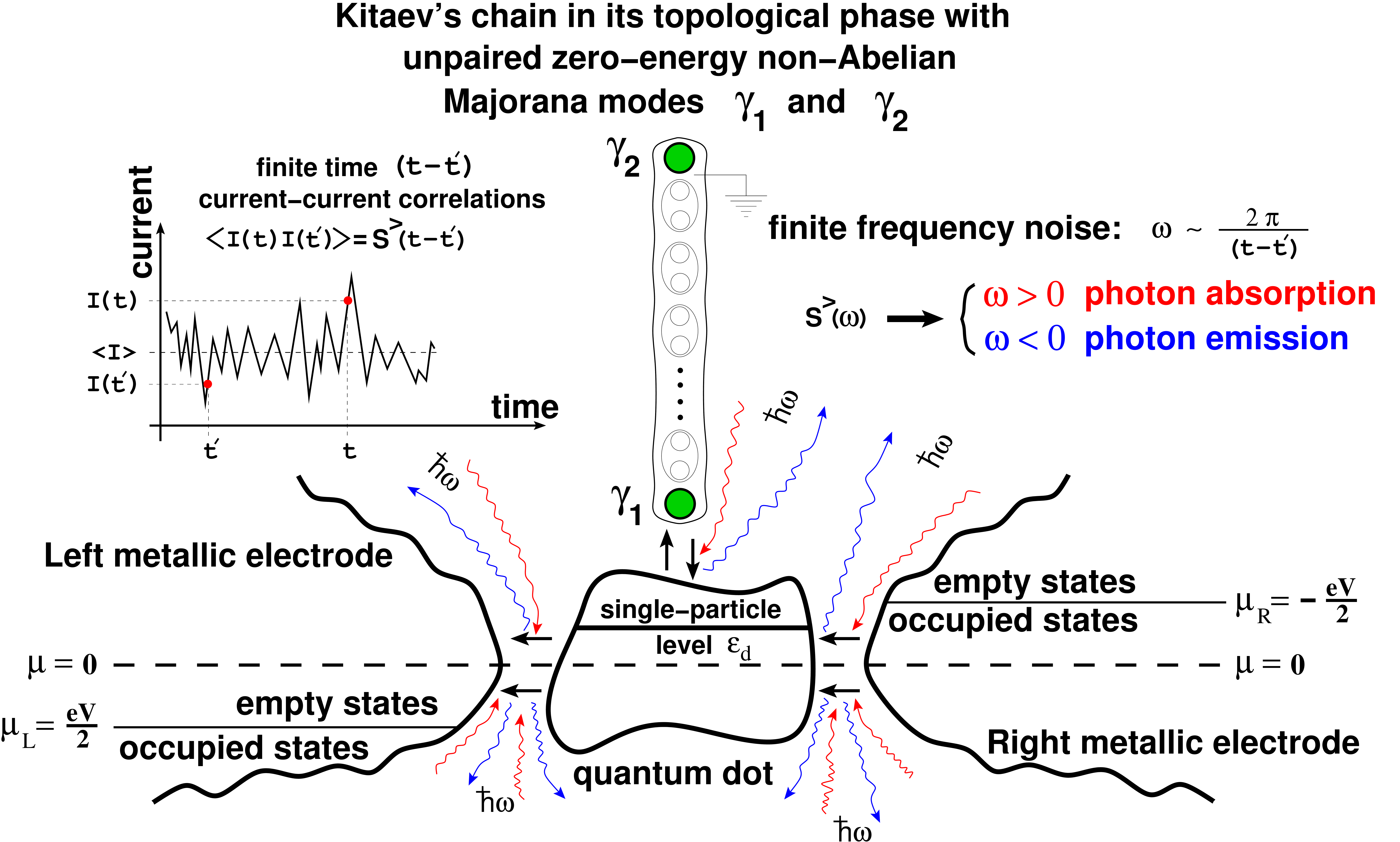}
\caption{\label{figure_1} A schematic illustration of a conceivable
  experimental setup. Quantum transport through a quantum dot is implemented
  via tunneling processes of two kinds: 1) tunneling of strength $\Gamma$
  between the quantum dot and normal metallic electrodes (left (L) and right
  (R) contacts); 2) tunneling of strength $|\eta|$ between the quantum dot and
  a topological superconductor implementing the Kitaev's chain with two
  Majorana bound states at its ends. Energetically both kinds of tunneling
  admit photon absorption and emission processes. One may access the photon
  absorption/emission spectra via the greater noise measured, {\it e.g.}, in
  the left contact, $S^>(t-t')=\langle I_L(t)I_L(t')\rangle$. The values of
  the electric current, $I(t)$ and $I(t')$, observed, respectively, at
  instants of time $t$ and $t'$, are correlated. Correlations at finite time
  intervals, $0<|t-t'|<\infty$, lead to noise $S^>(\omega)$ at finite
  frequencies $\omega$ providing absorption, $\omega>0$, and emission,
  $\omega<0$, spectra. Characteristic photon frequencies, $\omega_c$, are
  determined by the tunneling couplings,
  $\omega_c\sim\Gamma/\hbar,\,|\eta|/\hbar$. Applying a bias voltage $V$ to the
  contacts, one disturbs the system from equilibrium. The nonequilibrium noise
  is enriched by a new characteristic frequency, $\omega_c\sim |eV|/\hbar$.}
\end{figure}

In contrast to the static limit, quantum noise at finite frequencies in
Majorana setups is not so well investigated, especially in quantum dot setups,
such as, {\it e.g.}, the one shown in Fig. \ref{figure_1}, which admit a gate
tuning and may be used for simple tests of universal Majorana signatures. For
example, in topological superconductor-normal metal structures
\cite{Valentini_2016,Bathellier_2019} finite frequency noise induced by
Majorana zero modes has been investigated. However, universal fingerprints of
the Majorana finite frequency quantum noise have never been
discussed. Moreover, previously published results study noise itself and not
differential noise which is better (see below) suited to explore universal
nature of Majorana finite frequency noise. While the symmetrized finite
frequency noise may be interpreted as a classical stochastic force, the
nonsymmetrized, or quantum, finite-frequency noise admits interpretation
\cite{Clerk_2010} in terms of photon absorption and emission processes (shown
schematically in Fig. \ref{figure_1} by, respectively, red and blue wavy
arrows). However, in systems, where one can reach the quantum limit in a
realistic experiment, photon absorption/emission spectra, induced by Majorana
zero modes, remain unexplored.

Here, we focus on the differential quantum noise which is the derivative of
the quantum noise with respect to the frequency. This quantity has two
advantages in comparison with the quantum noise itself. First, the
differential quantum noise has universal units of $e^2$. Thus this quantity is
a natural tool to characterize universal features in Majorana finite frequency
quantum noise. Second, the differential quantum noise allows one to reveal in
high detail specific behavior of the quantum noise including possible fine
structures which one might easily skip when observing behavior of the quantum
noise itself. In particular, we analyze the photon absorption spectra at all
energy scales and reveal several remarkable properties of the differential
quantum noise as a function of the frequency $\nu$ ($h\nu=\hbar\omega$) at
very low temperatures $T$ (the lowest energy scale) and various bias voltages
$V$. We find that at low bias voltages 1) it has two universal plateaus,
$7e^2/4$ and $2e^2$; 2) at $h\nu=|eV|/2$ there is a very narrow antiresonance
with the full width at half of its minimum equal to $4k_\text{B}T$; 3) at
$h\nu=2|\eta|$ there is a resonance with full width at half of its maximum
equal to $\Gamma$; at large bias voltages 4) the antiresonance located at
$h\nu=|eV|/2$ turns into a resonance with the universal maximum $e^2/2$ and
the full width at half of the maximum equal to $\Gamma$; 5) at
$h\nu=2|\eta|\mp |eV|/2$ there develop two new resonances with the universal
maximum $e^2$ and the full width at half of the maximum equal to $\Gamma/2$;
6) the resonance located at $h\nu=2|\eta|$ turns into an
antiresonance-resonance pair in which the full widths of the antiresonance and
resonance at half of the minimum and maximum, respectively, are both equal to
$\Gamma$; 7) at all other frequencies the differential quantum noise
vanishes.

The paper is organized as follows. In Section \ref{phys_mod} we describe a
physical setup which can be implemented in realistic experiments.  The results
obtained for quantum finite frequency noise spectra are presented in Section
\ref{res}. With Section \ref{concl} we conclude the paper.
\section{Physical model}\label{phys_mod}
In this paper we present results for finite frequency quantum noise induced
by Majorana zero modes in the physical setup shown in Fig. \ref{figure_1}. The
system in this setup represents a quantum dot which is coupled to two massive
normal metals playing the role of the left and right contacts. The setup also
involves a grounded topological superconductor implementing the topological
phase of the Kitaev's model with two Majorana bound states at its ends one of
which is coupled to the quantum dot. As a result, the quantum dot states are
entangled with the Majorana degrees of freedom. When a bias voltage is applied
to the normal metals, electric current flows through this Majorana quantum dot
system and carries Majorana signatures both in its mean value and its
fluctuations. To understand the impact of the Majorana bound states on
fluctuations of the electric current, below we describe this setup more
precisely by specifying its full Hamiltonian which is the sum of the
Hamiltonians of the quantum dot, contacts, topological superconductor as well
as the Hamiltonians describing the interactions of the quantum dot with the
contacts and topological superconductor.

The quantum dot with one nondegenerate single-particle level $\epsilon_d$ is
described by the noninteracting Hamiltonian
\begin{equation}
\hat{H}_d=\epsilon_dd^\dagger d.
\label{Ham_QD}
\end{equation}
A gate voltage may shift $\epsilon_d$ relative to the chemical potential
$\mu$. The left and right contacts are normal metals with continuous energy
spectra $\epsilon_k$ characterized by a constant density of states
$\nu_c/2$. Their Hamiltonian is
\begin{equation}
\hat{H}_c=\sum_{l=L,R}\sum_k\epsilon_kc_{lk}^\dagger c_{lk}.
\label{Ham_cont}
\end{equation}
They are in an equilibrium state characterized by temperature $T$. An external
bias voltage $V$ may be applied to the contacts setting their chemical
potentials as $\mu_{L,R}=\pm eV/2$ (see Fig. \ref{figure_1} for $eV<0$). The
low-energy sector of the topological superconductor is modeled by
\begin{equation}
\hat{H}_{tsc}=i\xi\gamma_2\gamma_1/2,
\label{Ham_TSC}
\end{equation}
where $\gamma_{1,2}$ are the Majorana operators such that
$\gamma_{1,2}^\dagger=\gamma_{1,2}$. The anticommutation relations
$\{\gamma_i,\gamma_j\}=2\delta_{ij}$ establish the algebraic structure known
as the Clifford algebra. In general Majorana bound states have a finite
spatial overlap which we take into account via the overlap energy $\xi$. The
quantum dot interacts via tunneling (shown in Fig. \ref{figure_1} by black
arrows) with the contacts,
\begin{equation}
\hat{H}_{d-c}=\sum_{l=L,R}\sum_k\mathcal{T}c_{lk}^\dagger d+\text{H.c.},
\label{Ham_tun_D_C}
\end{equation}
and the topological superconductor,
\begin{equation}
\hat{H}_{d-tsc}=\eta^*d^\dagger\gamma_1+\text{H.c.}.
\label{Ham_tun_D_TSC}
\end{equation}
The corresponding tunneling strengths are $\Gamma=2\pi\nu_c|\mathcal{T}|^2$
and $|\eta|$.

To explore the quantum noise in a general nonequilibrium setup it is
convenient to employ the Keldysh generating functional \cite{Altland_2010},
\begin{equation}
\begin{split}
&Z[J_l(t)]=\int\mathcal{D}[\bar{\theta}(t),\theta(t)]e^{\frac{i}{\hbar}S_K[\bar{\theta}(t),\theta(t);J_l(t)]},\\
&\mathcal{D}[\bar{\theta}(t),\theta(t)]=\\
&=\mathcal{D}[\bar{\psi}(t),\psi(t)]\mathcal{D}[\bar{\phi}_{lk}(t),\phi_{lk}(t)]\mathcal{D}[\bar{\zeta}(t),\zeta(t)],
\end{split}
\label{Keld_gen_func}
\end{equation}
where $S_K[\bar{\theta}(t),\theta(t);J_l(t)]$ is the total Keldysh action with
the Grassmann fields of the quantum dot $(\bar{\psi}(t),\psi(t))$, contacts
$(\bar{\phi}_{lk}(t),\phi_{lk}(t))$ and the topological superconductor
$(\bar{\zeta}(t),\zeta(t))$ defined on the Keldysh closed time contour,
$t\in\mathcal{C}_K$. For $S_K[\bar{\theta}(t),\theta(t);J_l(t)]$ one has the
sum,
\begin{equation}
\begin{split}
&S_K[\bar{\theta}(t),\theta(t);J_l(t)]=\\
&=S_d[\bar{\psi}(t),\psi(t)]+S_c[\bar{\phi}_{lk}(t),\phi_{lk}(t)]+\\
&+S_{tsc}[\bar{\zeta}(t),\zeta(t)]+S_{d-c}[\bar{\psi}(t),\psi(t);\bar{\phi}_{lk}(t),\phi_{lk}(t)]+\\
&+S_{d-tsc}[\bar{\psi}(t),\psi(t);\bar{\zeta}(t),\zeta(t)]+\\
&+S_{scr}[\bar{\psi}(t),\psi(t);\bar{\phi}_{lk}(t),\phi_{lk}(t);J_l(t)],
\end{split}
\label{Tot_Keld_act}
\end{equation}
of the Keldysh actions describing the quantum dot, $S_d$, contacts, $S_c$,
topological superconductor, $S_{tsc}$, tunneling between the quantum dot and
contacts, $S_{d-c}$, tunneling between the quantum dot and topological
superconductor, $S_{d-tsc}$, and the source action $S_{scr}$.

As usual, the Keldysh actions $S_d$, $S_c$ and $S_{tsc}$ are represented in
the retarded-advanced space by upper triangular $2\times 2$ matrices whose
upper/lower diagonal elements are the corresponding inverse retarded/advanced
Green's functions and the upper off-diagonal elements are related to the
corresponding Fermi-Dirac distribution functions $f$ as $i\delta(1-2f)$,
$\delta\rightarrow 0^+$.

The tunneling actions,
\begin{equation}
\begin{split}
&S_{d-c}[\bar{\psi}(t),\psi(t);\bar{\phi}_{lk}(t),\phi_{lk}(t)]=\\
&=-\int_{-\infty}^\infty dt\sum_{l=L,R}\sum_{k}\bigl\{\mathcal{T}[\bar{\phi}_{lk+}(t)\psi_+(t)-\\
&-\bar{\phi}_{lk-}(t)\psi_-(t)]+\\
&+\mathcal{T}^*[\bar{\psi}_+(t)\phi_{lk+}(t)-\bar{\psi}_-(t)\phi_{lk-}(t)]\bigl\},
\end{split}
\label{Tunn_act_dc}
\end{equation}
\begin{equation}
\begin{split}
&S_{d-tsc}[\bar{\psi}(t),\psi(t);\bar{\zeta}(t),\zeta(t)]=\\
&=-\int_{-\infty}^\infty dt\bigl\{\eta^*[\bar{\psi}_+(t)\zeta_+(t)+\bar{\psi}_+(t)\bar{\zeta}_+(t)-\\
&-\bar{\psi}_-(t)\zeta_-(t)-\bar{\psi}_-(t)\bar{\zeta}_-(t)]+\\
&+\eta[\bar{\zeta}_+(t)\psi_+(t)+\zeta_+(t)\psi_+(t)-\\
&-\bar{\zeta}_-(t)\psi_-(t)-\zeta_-(t)\psi_-(t)]\bigl\},
\end{split}
\label{Tunn_act_ds}
\end{equation}
written on the forward/backward (subindex $+/-$) branches of $\mathcal{C}_K$,
take into account the interactions between the quantum dot and contacts,
Eq. (\ref{Tunn_act_dc}), as well as between the quantum dot and  topological
superconductor, Eq. (\ref{Tunn_act_ds}). Here the Grassmann fields
$\zeta_\pm(t)$ and $\bar{\zeta}_\pm(t)$ correspond to the complex Dirac
fermion representation of the Majorana operators. This representation clearly
shows how superconductivity appears in the formalism involving the anomalous
terms in Eq. (\ref{Tunn_act_ds}), $\bar{\psi}_\pm(t)\bar{\zeta}_\pm(t)$ and
$\zeta_\pm(t)\psi_\pm(t)$, which lead, as discussed below, to anomalous
contributions to the quantum noise. Note, that these anomalous contributions
result exclusively from the presence of the topological superconductor: they
could not have appeared if one had replaced the topological superconductor
with a normal system, {\it e.g.}, with another normal quantum dot.

Current-current correlators may be obtained via proper differentiation of
$Z[J_l(t)]$ with respect to the source field $J_l(t)$ coupled to the electric
current in the source action,
\begin{equation}
\begin{split}
&S_{scr}=-\int_{-\infty}^\infty dt\sum_{l=L,R}\sum_{q=+,-}J_{lq}(t)I_{lq}(t),\\
&I_{lq}(t)=\frac{ie}{\hbar}\sum_k[\mathcal{T}\bar{\phi}_{lkq}(t)\psi_q(t)-\mathcal{T}^*\bar{\psi}_q(t)\phi_{lkq}(t)],
\end{split}
\label{Scr_act}
\end{equation}
where the subscript $q$ denotes the forward ($q=+$) and backward ($q=-$)
branches of $\mathcal{C}_K$. In particular, for the correlations between the
values of the electric current in contacts $l$ and $l'$ taken at instants of
time $t$ and $t'$, $I_l(t)$ and $I_{l'}(t')$, one obtains
\begin{equation}
\begin{split}
&\langle I_l(t)I_{l'}(t')\rangle\equiv\langle I_{l-}(t)I_{l'+}(t')\rangle_{S_K}=\\
&=(i\hbar)^2\frac{\delta^2Z[J_l(t)]}{\delta J_{l-}(t)\delta J_{l'+}(t')}\biggl|_{J_{lq}(t)=0},
\end{split}
\label{Curr_curr_corr}
\end{equation}
where $\langle\ldots\rangle_{S_K}$ is the functional averaging with respect to
the total Keldysh action at vanishing sources,
\begin{equation}
\begin{split}
&\langle\mathcal{F}[\bar{\theta}(t),\theta(t)]\rangle_{S_K}\equiv\\
&\equiv\int\mathcal{D}[\bar{\theta}(\tilde{t}),\theta(\tilde{t})]e^{\frac{\mathrm{i}}{\hbar}S_K[\bar{\theta}(\tilde{t}),\theta(\tilde{t})]}\mathcal{F}[\bar{\theta}(t),\theta(t)],
\end{split}
\label{Aver_zero_scr_fld}
\end{equation}
\begin{equation}
\begin{split}
&S_K[\bar{\theta}(t),\theta(t)]\equiv S_K[\bar{\theta}(t),\theta(t);J_l(t)=0]=\\
&=S_d[\bar{\psi}(t),\psi(t)]+S_c[\bar{\phi}_{lk}(t),\phi_{lk}(t)]+\\
&+S_{tsc}[\bar{\zeta}(t),\zeta(t)]+S_{d-c}[\bar{\psi}(t),\psi(t);\bar{\phi}_{lk}(t),\phi_{lk}(t)]+\\
&+S_{d-tsc}[\bar{\psi}(t),\psi(t);\bar{\zeta}(t),\zeta(t)].
\end{split}
\label{Tot_Keld_act_zero_scr_fld}
\end{equation}

Integrating first out the contacts Grassmann fields and afterwards taking the
second derivative in Eq. (\ref{Curr_curr_corr}) it is easy to see that the
current-current correlator is the sum,
\begin{equation}
\langle I_l(t)I_{l'}(t')\rangle=S_{1_{ll'}}(t,t')+S_{2_{ll'}}(t,t'),
\label{Curr_curr_corr_sum_1_2}
\end{equation}
of a one-particle term $S_{1_{ll'}}(t,t')$ and also a two-particle term
$S_{2_{ll'}}(t,t')$.

\begin{widetext}
Restoring the field integrals over the contacts Grassmann fields one can
express the terms  $S_{1_{ll'}}(t,t')$ and $S_{2_{ll'}}(t,t')$ using the
functional averaging defined in Eqs. (\ref{Aver_zero_scr_fld}) and
(\ref{Tot_Keld_act_zero_scr_fld}),
\begin{equation}
\begin{split}
&S_{1_{ll'}}(t,t')=\delta_{ll'}|\mathcal{T}|^2(-i)\biggl(\frac{e}{2\hbar}\biggl)^2
\sum_k\bigl\{\bigl\langle[\psi_1(t')+\psi_2(t')][-\bar{\psi}_1(t)+\bar{\psi}_2(t)]\bigl\rangle_{S_K}
\bigl[G_{lk}^R(t-t')+G_{lk}^K(t-t')-\\
&-G_{lk}^A(t-t')\bigl]+
\bigl\langle[\psi_1(t)-\psi_2(t)][\bar{\psi}_1(t')+\bar{\psi}_2(t')]\bigl\rangle_{S_K}\bigl[-G_{lk}^R(t'-t)+G_{lk}^K(t'-t)+G_{lk}^A(t'-t)\bigl]\bigl\},
\end{split}
\label{Curr_curr_corr_1}
\end{equation}
\begin{equation}
\begin{split}
&S_{2_{ll'}}(t,t')=|\mathcal{T}|^4\frac{(-1)}{\hbar^2}\biggl(\frac{e}{2\hbar}\biggl)^2\int_{-\infty}^{\infty}dt_1\int_{-\infty}^{\infty}dt_2\sum_{k_1,k_2}
\biggl\langle\bigl\{\bigl[-\psi_1(t)+\psi_2(t)\bigl]
\bigl[\bar{\psi}_1(t_2)G_{lk_2}^R(t_2-t)-\\
&-\bar{\psi}_1(t_2)G_{lk_2}^K(t_2-t)-\bar{\psi}_2(t_2)G_{lk_2}^A(t_2-t)\bigl]-
\bigl[G_{lk_2}^R(t-t_2)\psi_1(t_2)+G_{lk_2}^K(t-t_2)\psi_2(t_2)-\\
&-G_{lk_2}^A(t-t_2)\psi_2(t_2)\bigl]\bigl[-\bar{\psi}_1(t)+\bar{\psi}_2(t)\bigl]\bigl\}
\bigl\{\bigl[\psi_1(t')+\psi_2(t')\bigl]\bigl[\bar{\psi}_1(t_1)G_{l'k_1}^R(t_1-t')+\bar{\psi}_1(t_1)G_{l'k_1}^K(t_1-t')+\\
&+\bar{\psi}_2(t_1)G_{l'k_1}^A(t_1-t')\bigl]-\bigl[G_{l'k_1}^R(t'-t_1)\psi_1(t_1)+G_{l'k_1}^K(t'-t_1)\psi_2(t_1)+\\
&+G_{l'k_1}^A(t'-t_1)\psi_2(t_1)\bigl]\bigl[\bar{\psi}_1(t')+\bar{\psi}_2(t')\bigl]\bigl\}\biggl\rangle_{S_K},
\end{split}
\label{Curr_curr_corr_2}
\end{equation}
where the Keldysh rotation,
\begin{equation}
\begin{split}
&\psi_q(t)=\frac{1}{\sqrt{2}}\bigl[\psi_1(t)+q\psi_2(t)\bigl],\quad \bar{\psi}_q(t)=\frac{1}{\sqrt{2}}\bigl[\bar{\psi}_2(t)+q\bar{\psi}_1(t)\bigl],
\end{split}
\label{Keld_rot_qd}
\end{equation}
has been used and $G_{lk}^{R,A,K}(t-t')$ are the conventional contacts
retarded, advanced and Keldysh Green's functions, respectively.
\end{widetext}

The one-particle contribution in Eq. (\ref{Curr_curr_corr_1}) involves
averages of products of only two Grassmann fields of the quantum dot. This
contribution contains only normal terms,
$\langle\psi_r(t)\bar{\psi}_{r'}(t')\rangle_{S_K}$, ($r,r'=1,2$). Obviously,
the anomalous terms, $\langle\psi_r(t)\psi_{r'}(t')\rangle_{S_K}$ or
$\langle\bar{\psi}_r(t)\bar{\psi}_{r'}(t')\rangle_{S_K}$, cannot appear in
$S_{1_{ll'}}(t,t')$.

The anomalous terms definitely arise when one averages products of four
Grassmann fields in the two-particle contribution,
Eq. (\ref{Curr_curr_corr_2}). Since the total Keldysh action in
Eq. (\ref{Tot_Keld_act_zero_scr_fld}) is quadratic, the average of a product
of four Grassmann fields can be written as a sum of products of the averages
of two Grassmann fields according to Wick's theorem. In particular,
\begin{equation}
\begin{split}
&\langle\psi^1_{r_1}\bar{\psi}^2_{r_2}\psi^3_{r_3}\bar{\psi}^4_{r_4}\rangle_{S_K}=\langle\psi^1_{r_1}\bar{\psi}^2_{r_2}\rangle_{S_K}\langle\psi^3_{r_3}\bar{\psi}^4_{r_4}\rangle_{S_K}-\\
&-\!\langle\psi^1_{r_1}\bar{\psi}^4_{r_4}\rangle_{S_K}\langle\psi^3_{r_3}\bar{\psi}^2_{r_2}\rangle_{S_K}\!-\!\langle\psi^1_{r_1}\psi^3_{r_3}\rangle_{S_K}\langle\bar{\psi}^2_{r_2}\bar{\psi}^4_{r_4}\rangle_{S_K}.
\end{split}
\label{Avr_4_Grass_fld}
\end{equation}
In Eq. (\ref{Avr_4_Grass_fld}) $\psi^1_{r_1}$, $\bar{\psi}^2_{r_2}$,
$\psi^3_{r_3}$, $\bar{\psi}^4_{r_4}$ ($r_{1,2,3,4}=1,2$) schematically denote
four Grassmann fields in a particular contribution resulting from a product of
four square brackets in the integrand of (\ref{Curr_curr_corr_2}). The
superscript enumerates, from left to right, those four square brackets from
which the corresponding Grassmann fields are taken to form the particular
product (\ref{Avr_4_Grass_fld}). The subscript provides the corresponding
component of the retarded-advanced space.

All those terms in Eq. (\ref{Curr_curr_corr_2}) which correspond to the third
term in the right hand side of Eq. (\ref{Avr_4_Grass_fld}) give the anomalous
two-particle contributions to $\langle I_l(t)I_{l'}(t')\rangle$. We emphasize
once again that the appearance of these anomalous contributions in the
mathematical formalism is a direct consequence of superconductivity present in
the physical system. The anomalous contributions would have never appeared if
one had replaced the topological superconductor with, {\it e.g.}, another
normal quantum dot.

Let us introduce a "particle-hole" space via the Grassmann fields $\psi_{ar}$
($a=p,h$; $r=1,2$),
\begin{equation}
\begin{split}
&\psi_{ar}(t)\equiv\bar{\psi}_r(t),\quad a=p,\\
&\psi_{ar}(t)\equiv\psi_r(t),\quad a=h.
\end{split}
\label{P_h_space}
\end{equation}
Then the averages in Eq. (\ref{Curr_curr_corr_1}) as well as the averages
appearing in Eq. (\ref{Curr_curr_corr_2}) via Eq. (\ref{Avr_4_Grass_fld}) are
expressed in terms of the hole-particle, hole-hole and particle-particle
retarded, advanced and Keldysh Green's functions of the quantum dot:
\begin{widetext}
\begin{equation}
\langle\psi_{hr}(t)\psi_{pr'}(t')\rangle_{S_K}=
\begin{pmatrix}
i\,\mathcal{G}_{hp}^R(t-t') & i\,\mathcal{G}_{hp}^K(t-t') \\
0 & i\,\mathcal{G}_{hp}^A(t-t')
\end{pmatrix},
\label{G_f_hp}
\end{equation}
\begin{equation}
\langle\psi_{hs}(t)\psi_{hs'}(t')\rangle_{S_K}=
\begin{pmatrix}
i\,\mathcal{G}_{hh}^K(t-t') & i\,\mathcal{G}_{hh}^R(t-t') \\
i\,\mathcal{G}_{hh}^A(t-t') & 0
\end{pmatrix},
\label{G_f_hh}
\end{equation}
\begin{equation}
\langle\psi_{ps}(t)\psi_{ps'}(t')\rangle_{S_K}=
\begin{pmatrix}
0 & i\,\mathcal{G}_{pp}^A(t-t') \\
i\,\mathcal{G}_{pp}^R(t-t') & i\,\mathcal{G}_{pp}^K(t-t')
\end{pmatrix}.
\label{G_f_pp}
\end{equation}
\end{widetext}

The total Keldysh action, Eq. (\ref{Tot_Keld_act_zero_scr_fld}), is quadratic
and thus the quantum dot Green's functions are the elements of the inverse
kernel of the effective Keldysh action obtained from the total Keldysh action
after integrating out the Grassmann fields of the contacts and topological
superconductor. Although the effective Keldysh action is nonlocal in time, one
can transform to the frequency domain and easily find the expressions for the
hole-particle, hole-hole and particle-particle retarded, advanced
\cite{Smirnov_2015} and Keldysh \cite{Smirnov_2018} Green's functions of the
quantum dot:
\begin{widetext}
\begin{equation}
\begin{split}
&\mathcal{G}^R_{hp}(\epsilon)=\frac{N^R_{hp}(\epsilon)}{f(\epsilon)},\quad
\mathcal{G}^R_{hh}(\epsilon)=\frac{-8\hbar(\eta^*)^2\epsilon}{f(\epsilon)},\quad \mathcal{G}^R_{pp}(\epsilon)=\frac{-8\hbar\eta^2\epsilon}{f(\epsilon)},\\
&\mathcal{G}^A_{hp}(\epsilon)=[\mathcal{G}^R_{hp}(\epsilon)]^*,\quad \mathcal{G}^A_{hh}(\epsilon)=[\mathcal{G}^R_{pp}(\epsilon)]^*,\quad\mathcal{G}^A_{pp}(\epsilon)=[\mathcal{G}^R_{hh}(\epsilon)]^*,\\
&f(\epsilon)=4\epsilon^4-\epsilon^2(\Gamma^2+4\epsilon_d^2+4\xi^2+16|\eta|^2)+\xi^2(\Gamma^2+4\epsilon_d^2)+i\,4\Gamma[\epsilon^3-\epsilon(\xi^2+2|\eta|^2)],\\
&N^R_{hp}(\epsilon)=2\hbar\{-4|\eta|^2\epsilon-(\xi^2-\epsilon^2)[i\Gamma+2(\epsilon_d+\epsilon)]\},
\end{split}
\label{R_A_G_f}
\end{equation}
\begin{equation}
\begin{split}
&\mathcal{G}^K_{hp}(\epsilon)=\frac{N^K_{hp}(\epsilon)}{|f(\epsilon)|^2},\quad
\mathcal{G}^K_{hh}(\epsilon)=\frac{N^K_{hh}(\epsilon)}{|f(\epsilon)|^2},\quad\mathcal{G}^K_{pp}(\epsilon)=\frac{N^K_{pp}(\epsilon)}{|f(\epsilon)|^2},\\
&N^K_{hp}(\epsilon)=-2\,i\,\Gamma\hbar\bigl\{(F_L(\epsilon)+F_R(\epsilon))(\xi^2-\epsilon^2)^2[\Gamma^2+4(\epsilon_d+\epsilon)^2]+\\
&+16|\eta|^2\epsilon[(F_L(\epsilon)+F_R(\epsilon))(\xi^2-\epsilon^2)(\epsilon_d+\epsilon)+|\eta|^2\epsilon(F_L(\epsilon)-F_L(-\epsilon)+F_R(\epsilon)-F_R(-\epsilon))]\bigl\},\\
&N^K_{hh}(\epsilon)=8\Gamma\hbar(\eta^*)^2\epsilon\bigl\{(\xi^2-\epsilon^2)[(F_L(\epsilon)+F_L(-\epsilon)+F_R(\epsilon)+F_R(-\epsilon))(\Gamma-2\,i\,\epsilon_d)+\\
&+2\,i\,\epsilon(F_L(-\epsilon)-F_L(\epsilon)+F_R(-\epsilon)-F_R(\epsilon))]+4\,i\,|\eta|^2\epsilon(F_L(-\epsilon)-F_L(\epsilon)+F_R(-\epsilon)-F_R(\epsilon))\bigl\},\\
&N^K_{pp}(\epsilon)=-(N^K_{hh}(\epsilon))^*,
\end{split}
\label{K_G_f}
\end{equation}
\end{widetext}
where $F_{L,R}(\epsilon)$ are defined through the Fermi-Dirac distributions
$f_{L,R}(\epsilon)$ of, respectively, the left and right contacts as
$F_{L,R}(\epsilon)\equiv 1-2f_{L,R}(\epsilon)$.

Below we use Eqs. (\ref{Curr_curr_corr_1}), (\ref{Curr_curr_corr_2}),
(\ref{Avr_4_Grass_fld}) and (\ref{G_f_hp})-(\ref{K_G_f}) as a complete set of
expressions to obtain the Fourier transform of
$\langle I_l(t)I_{l'}(t')\rangle$ via numerical integrations. We perform these
numerical integrations in the frequency domain in order to make use of
Eqs. (\ref{R_A_G_f}) and (\ref{K_G_f}).

To be more precise, we investigate the greater current-current correlator in
the left contact, 
\begin{equation}
S^>(t-t';V)=\langle I_L(t)I_L(t')\rangle-I^2_L(V),
\label{Grt_noise_t}
\end{equation}
where $I_L(V)$ is the corresponding mean current. The Fourier transform
\begin{equation}
S^>(\omega,V)=\int_{-\infty}^\infty dt e^{i\omega t}S^>(t,V),
\label{Grt_noise_fr}
\end{equation}
provides the photon absorption/emission spectra $S^{ab/em}$,
\begin{equation}
S^{ab/em}(\omega,V)=
\begin{cases}
S^>(\omega,V),\quad\omega>0\\
S^>(\omega,V),\quad\omega<0
\end{cases},
\label{Ab_em_spec}
\end{equation}
which are of direct experimental interest.
\begin{figure}
\includegraphics[width=8.0 cm]{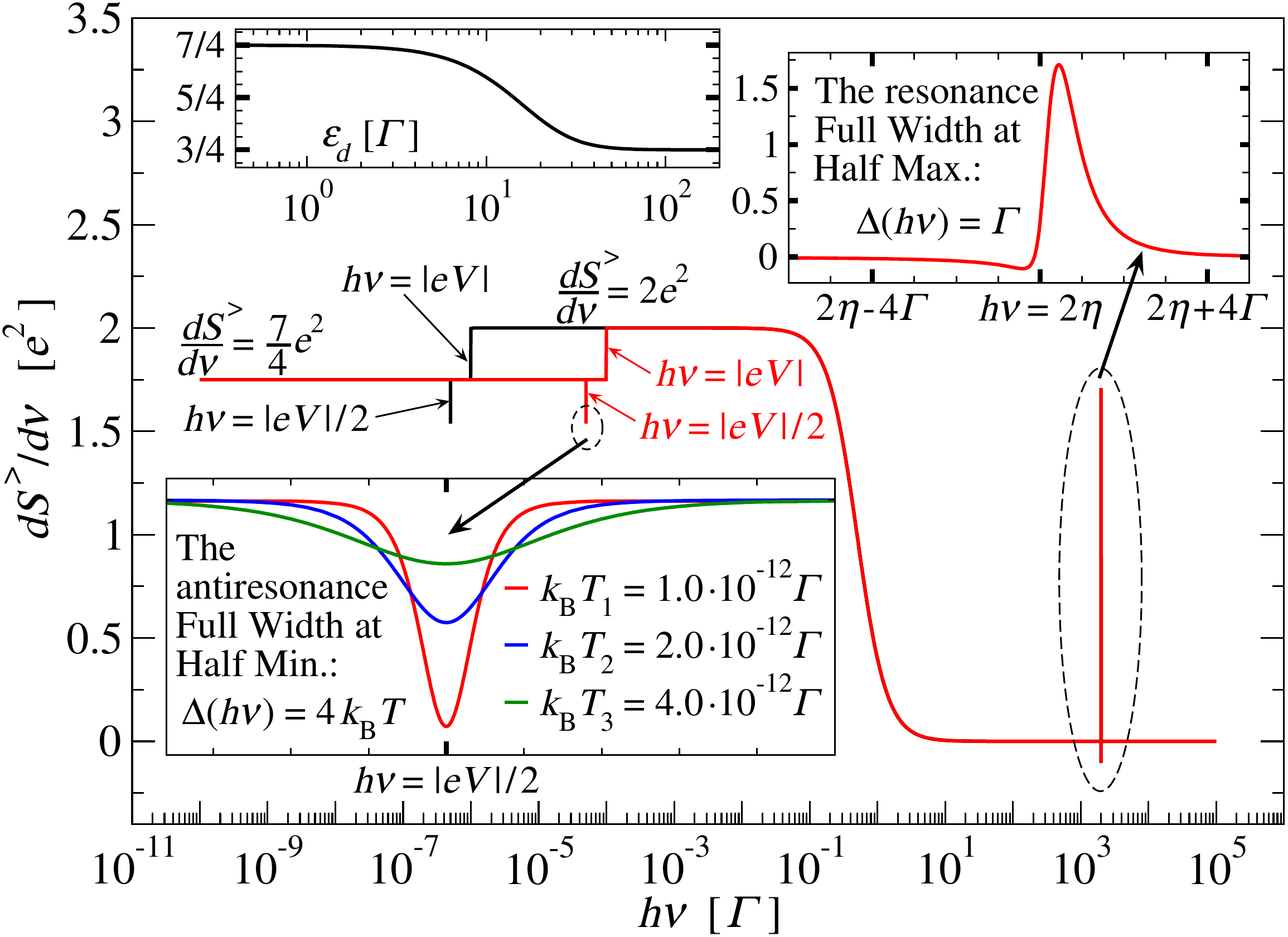}
\caption{\label{figure_2} The differential quantum noise
  $\partial S^>(\nu,V)/\partial\nu$ as a function of the frequency $\nu$
  ($h\nu=\hbar\omega$) is shown for the case of very small bias voltages,
  $|eV|\ll\Gamma$. Here $k_\text{B}T/\Gamma=10^{-12}$, $\epsilon_d/\Gamma=8.0$,
  $|\eta|/\Gamma=10^3$ and $\xi/\Gamma=10^{-4}$.The black and red curves are
  for $|eV|/\Gamma=10^{-6}$ and $|eV|/\Gamma=10^{-4}$, respectively. The lower
  inset shows the antiresonance whose minimum is located at $h\nu=|eV|/2$. The
  upper left inset shows the dependence of this minimum on $\epsilon_d$ which
  may be controlled by a gate voltage. The upper right inset shows the
  antiresonance-resonance pair, located in the vicinity of $h\nu=2|\eta|$.}
\end{figure}

To access finite frequency quantum noise induced by Majorana zero modes we
explore the transport regime governed by strong Majorana tunneling,
\begin{equation}
|\eta|>\text{max}\{|\epsilon_d|,|eV|,k_\text{B}T,\Gamma,\xi\}.
\label{M_tunn_reg}
\end{equation}
We obtain the differential quantum noise
$\partial S^>(\nu,V)/\partial\nu$ ($h\nu=\hbar\omega$) via, first, numerical
integrations (using, $e.g.$, Simpson's rule) in the frequency domain,
providing $S^>(\nu,V)$ on a fine frequency grid, and, afterwards, applying
finite differences (using, $e.g.$, forward difference or backward difference)
to obtain $\partial S^>(\nu,V)/\partial\nu$. Due to high precision in the
numerical frequency integrations necessary to provide highly precise values of
$S^>(\nu,V)$ to later obtain reliable values of the derivative
$\partial S^>(\nu,V)/\partial\nu$ the numerical computations are very time
consuming but still feasible. In particular, below we focus on the photon
absorption spectra, {\it i.e.} on $\partial S^>(\nu,V)/\partial\nu$ for
$\nu>0$.
\section{Results}\label{res}
In Fig. \ref{figure_2} we show the results for the differential quantum noise
in the regime $|eV|\ll\Gamma$. It reveals a highly detailed structure of the
photon absorption ($\nu>0$) spectrum. For frequencies $h\nu\ll\Gamma$ the
differential noise is represented by two plateaus with universal values
$\partial S^>(\nu,V)/\partial\nu=7e^2/4$ for $h\nu<|eV|$ and
$\partial S^>(\nu,V)/\partial\nu=2e^2$ for $h\nu>|eV|$. At $h\nu=|eV|/2$ the
differential noise shows an antiresonance having the full width equal to
$4k_\text{B}T$ at half of its minimum, as shown in the lower inset for
$|eV|/\Gamma=10^{-4}$ and three different temperatures,
$k_\text{B}T_1/\Gamma=10^{-12}$ (red),
$k_\text{B}T_2/\Gamma=2.0\cdot 10^{-12}$ (blue) and
$k_\text{B}T_3/\Gamma=4.0\cdot 10^{-12}$ (green). This antiresonance is not
universal, {\it i.e.} it depends on the gate voltage tuning $\epsilon_d$. The
minimum of the antiresonance as a function of $\epsilon_d$ is shown in the
upper left inset. It has two asymptotics. The minimal value, $3e^2/4$, is
reached at large values of $\epsilon_d$ while the maximal value, $7e^2/4$, is
reached in the limit $\epsilon_d\rightarrow 0$. Therefore, the antiresonance
disappears when $\epsilon_d=0$. Near $h\nu=2|\eta|$ the differential noise
exhibits a strong resonance having the full width equal to $\Gamma$ at half of
its maximum as shown in the upper right inset. Just before this strong
resonance one also notices a very shallow antiresonance. For the present case
of very small bias voltages, $|eV|\ll\Gamma$, this antiresonance-resonance
pair, located in the vicinity of $h\nu=2|\eta|$, is independent of the bias
voltage $V$.
\begin{figure}
\includegraphics[width=8.0 cm]{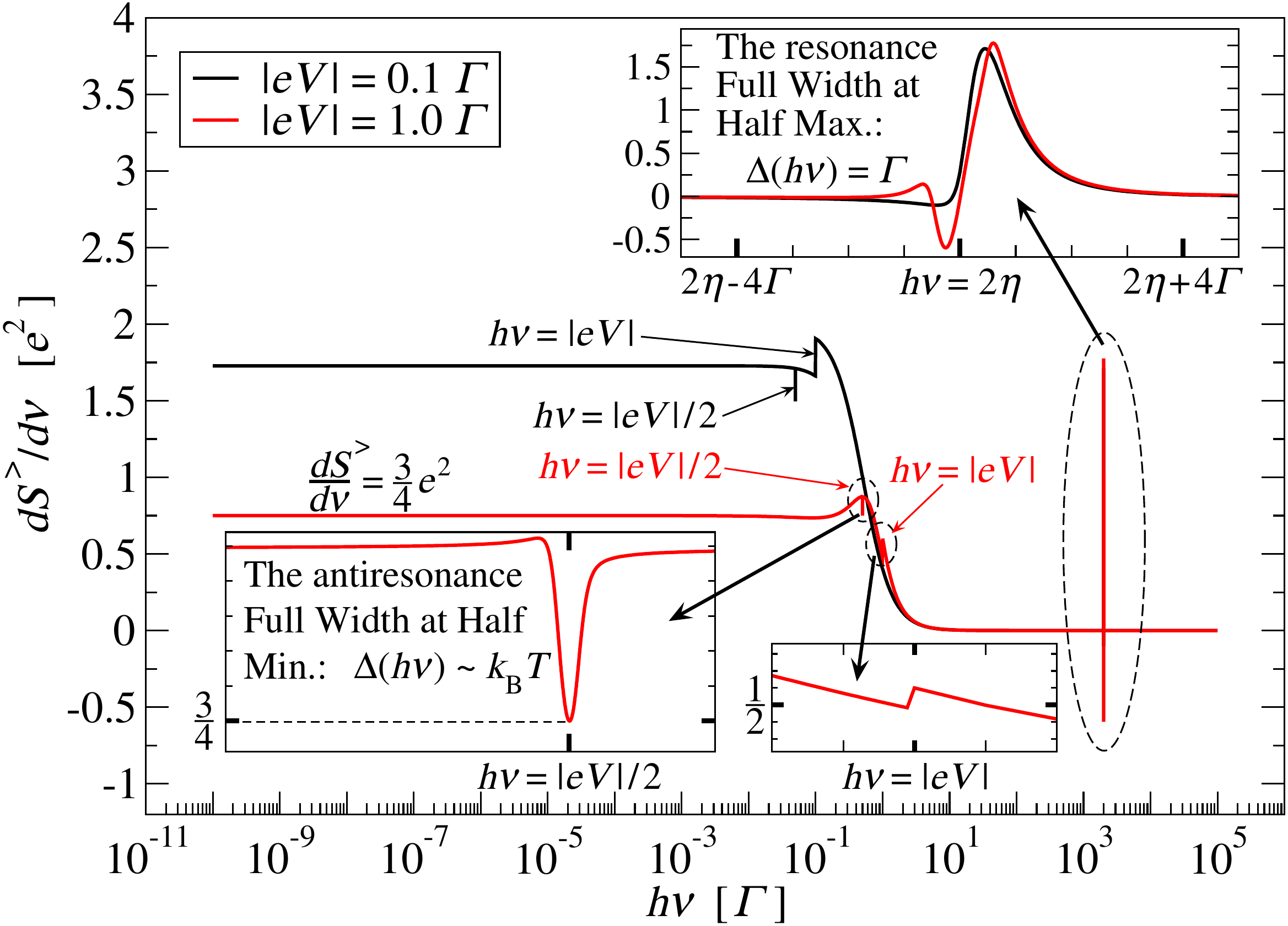}
\caption{\label{figure_3} The differential quantum noise
  $\partial S^>(\nu,V)/\partial\nu$ as a function of the frequency $\nu$ is
  shown for the case of small and moderate bias voltages. The black and red
  curves show $\partial S^>(\nu,V)/\partial\nu$ for $|eV|/\Gamma=0.1$ and
  $|eV|/\Gamma=1.0$, respectively. The other parameters are the same as in
  Fig. \ref{figure_2}. The lower left inset shows the antiresonance which is
  still present in the case $|eV|/\Gamma=1.0$ and has its minimum located at
  $h\nu=|eV|/2$. The lower right inset shows a jump at $h\nu=|eV|$ for the
  case $|eV|/\Gamma=1.0$. The upper inset shows for the case $|eV|/\Gamma=1.0$
  the resonance-antiresonance-resonance trio located in the vicinity of
  $h\nu=2|\eta|$.}
\end{figure}

In Fig. \ref{figure_3} the photon absorption spectrum is analyzed in the
regime $|eV|\lesssim\Gamma$, {\it i.e.}, respectively, for small bias
voltages, $|eV|<\Gamma$, and moderate bias voltages, $|eV|\sim\Gamma$. As one
can see, in this regime the differential noise has a plateau at
$h\nu\ll|eV|$. The value of this plateau is, however, suppressed below its
universal maximum $7e^2/4$ reached in the regime of very small bias voltages,
$|eV|\ll\Gamma$, for $h\nu<|eV|$ (see Fig. \ref{figure_2}). When
$|eV|=\Gamma$, the universal value of this plateau is $3e^2/4$ (red curve). As
in the regime $|eV|\ll\Gamma$, shown in Fig. \ref{figure_2}, also in the
present case, $|eV|\lesssim\Gamma$, the differential quantum noise has a jump
at $h\nu=|eV|$. However, now this jump is significantly reduced, as shown in
the lower right inset for $|eV|=\Gamma$, and, as one can also see, this is a
jump in a decreasing function and not a jump separating two universal plateaus
of the differential noise as is the case in Fig. \ref{figure_2}. At
$h\nu=|eV|/2$ the situation starts to qualitatively change when the bias
voltage grows. In particular, when $|eV|=\Gamma$ (red curve), there develops a
wide resonance whose full width is of order $\Gamma$ at half of its
maximum. This wide resonance is split in its very middle, {\it i.e.} at
$h\nu=|eV|/2$, by the extremely narrow antiresonance which has already
developed in the regime of very small bias voltages, $|eV|\ll\Gamma$, as shown
in the lower inset of Fig. \ref{figure_2}. Here, for the bias voltage
$|eV|=\Gamma$, the full width of this extremely narrow antiresonance is still
of order $k_\text{B}T$ at half of its minimum as shown in the lower left
inset. The minimum of this antiresonance is equal to the value of the plateau
at $h\nu\ll |eV|$, {\it i.e.} to $3e^2/4$, as emphasized by the horizontal
dashed line in the lower left inset. In the vicinity of $h\nu=2|\eta|$
qualitative changes in the behavior of the differential noise become visible
too. As one can see in the upper inset, when $|eV|=\Gamma$ (red curve), there
appears a new resonance just before the antiresonance-resonance pair which has
already developed in the regime of very small bias voltages, $|eV|\ll\Gamma$
as shown in the upper right inset of Fig. \ref{figure_2}. Also note that in
this resonance-antiresonance-resonance trio, located in the vicinity of
$h\nu=2|\eta|$, the antiresonance becomes stronger in comparison with its
predecessor antiresonance from the antiresonance-resonance pair shown in the
vicinity of $h\nu=2|\eta|$ in Fig. \ref{figure_2} (see its upper right inset)
in the regime of very small bias voltages, $|eV|\ll\Gamma$.
\begin{figure}
\includegraphics[width=8.0 cm]{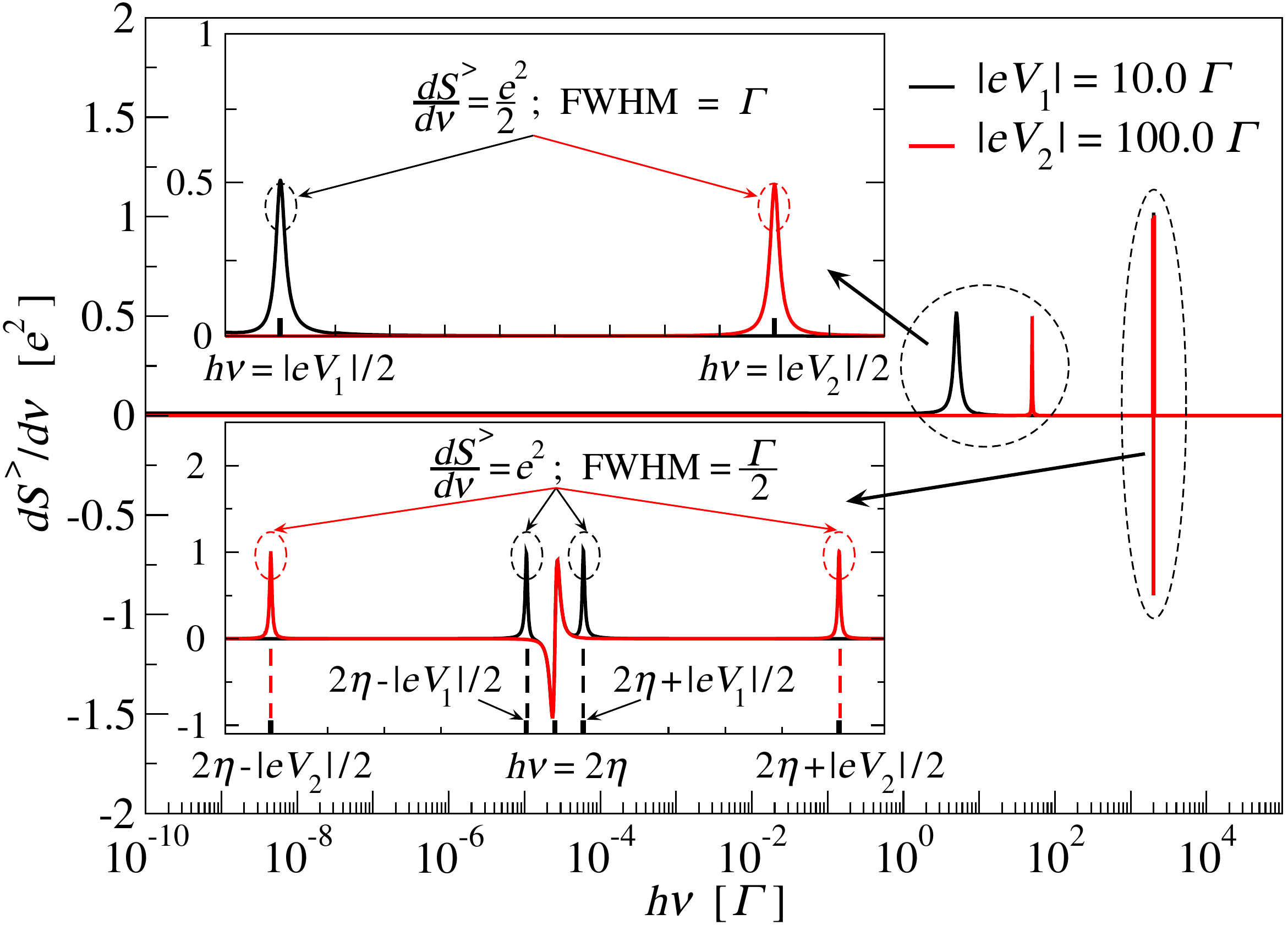}
\caption{\label{figure_4} The differential quantum noise
  $\partial S^>(\nu,V)/\partial\nu$ as a function of the frequency $\nu$ is
  shown for the case of large and very large bias voltages. The black and red
  curves show $\partial S^>(\nu,V)/\partial\nu$ for, respectively, $V=V_1$ and
  $V=V_2$ with $|eV_1|/\Gamma=10.0$ and $|eV_2|/\Gamma=100.0$. The other
  parameters are the same as in Fig. \ref{figure_2}. The upper inset shows the
  resonances at $h\nu=|eV_1|/2$ (black curve) and $h\nu=|eV_2|/2$ (red
  curve). The lower inset shows the antiresonance-resonance pair located in
  the vicinity of $h\nu=2|\eta|$ as well as the resonances located at
  $h\nu=2|\eta|\mp |eV_1|/2$ (black curve) and $h\nu=2|\eta|\mp |eV_2|/2$ (red
  curve).}
\end{figure}

Our further results, presented in Fig. \ref{figure_4}, show the differential
quantum noise in the regime of large, $|eV|\gtrsim\Gamma$, and very large,
$|eV|\gg\Gamma$, bias voltages. In this strongly nonequilibrium regime the
differential noise vanishes at all frequencies except for vicinities of a few
characteristic frequencies. First, the antiresonance located at $h\nu=|eV|/2$
in Figs. \ref{figure_2} and \ref{figure_3} is fully washed out in strong
nonequilibrium. Instead, at $h\nu=|eV|/2$ there develops a resonance shown in
the upper inset for $V=V_1$ (black curve) and $V=V_2$ (red curve). The
universal (independent of $\epsilon_d$) maximum of this resonance is equal to
$e^2/2$ while its full width at half of the maximum is equal to $\Gamma$ as
demonstrated in the upper inset. Further, in the vicinity of $h\nu=2|\eta|$
there develops an antiresonance-resonance pair shown in the lower inset. This
antiresonance-resonance pair is independent of the bias voltage $V$ for
$|eV|>\Gamma$. Finally, at frequencies $h\nu=2|\eta|\mp |eV|/2$ there develop
two resonances. The universal (independent of $\epsilon_d$) maximum of these
resonances is equal to $e^2$ while their full width at half of the maximum is
equal to $\Gamma/2$ as demonstrated in the lower inset for $V=V_1$ (black
curve) and $V=V_2$ (red curve).

The universal Majorana properties of finite frequency quantum noise presented
above may intuitively be explained as follows.

First of all, from Eq. (\ref{R_A_G_f}) one obtains the spectral function
$\nu(\epsilon)\equiv(-1/\pi\hbar)\text{Im}[\mathcal{G}^R_{hp}(\epsilon)]$.
The spectral function provides the quasiparticle spectrum of the quantum
dot. It is easy to find from $\nu(\epsilon)$ that for $\xi=0$ the
quasiparticle spectrum of the quantum dot consists of three peaks located at
$\epsilon=0,\mp 2|\eta|$. These peaks have the same maximum equal to
$1/(\pi\Gamma)$. The full width of the peak at $\epsilon=0$ is equal to
$\Gamma$ at half of its maximum. The full width of the peaks at
$\epsilon=\mp 2|\eta|$ is equal to $\Gamma/2$ at half of the
maximum. Moreover, for $\xi\neq 0$ there appears a dip located near
$\epsilon=0$. For $\epsilon_d\geqslant\Gamma$ the minimum of this dip vanishes
as $\nu(\epsilon)\sim\Gamma/(5\pi\epsilon_d^2)$. This dip is extremely narrow
for small $\xi$ and large $\eta$. Its full width at half of its minimum is
proportional to $[(\epsilon_d\xi/\eta)^2]/\Gamma$.

Now, having in mind the above structure of the spectral function, it is
obvious from the energy conservation that photon absorption leading to the
excitation of the quantum dot by energy $\Delta\epsilon=2|\eta|$ is
responsible for the antiresonance-resonance pair (see Fig. \ref{figure_2})
located at $h\nu=2|\eta|$ and for the two resonances (see Fig. \ref{figure_4})
located at $h\nu=2|\eta|\mp|eV|/2$. The latter two resonances also involve
tunneling processes after which the quasiparticle energy in the final state is
decreased (increased) by $\Delta\epsilon_{qp}=|eV|/2$, {\it i.e.} tunneling
processes {\it topological superconductor (left contact)} $\rightarrow$
{\it quantum dot} $\rightarrow$
{\it left contact (topological superconductor)}. Likewise, the energy  
conservation admits photon absorption with weak excitations of the quantum
dot, that is with excitation energies $\Delta\epsilon\lesssim\Gamma$, but
involve tunneling processes after which the quasiparticle energy in the final
state is increased by $\Delta\epsilon_{qp}=|eV|/2$,
{\it left contact} $\rightarrow$ {\it quantum dot} $\rightarrow$
{\it topological superconductor}. Such processes are responsible for the
formation of the antiresonance ($|eV|<\Gamma$) or resonance ($|eV|\gg\Gamma$)
at $h\nu=|eV|/2$. Here, the above discussed dip in the spectral function near
$\epsilon=0$ is detected at small bias voltages and temperatures leading to
the antiresonance (see Fig. \ref{figure_2}) in the differential noise at
$h\nu=|eV|/2$. However, for large bias voltages this extremely narrow dip in
the spectral function is not any more detected and thus effectively one has
the situation where $\xi=0$. Therefore, the resonance of the spectral function
located around $\epsilon=0$ leads instead of the antiresonance to the
resonance (see Fig. \ref{figure_4}) in the differential noise at
$h\nu=|eV|/2$. The full width of this resonance at half of its maximum is
equal to $\Gamma$. This exactly corresponds to the full width of the spectral
function peak at half of its maximum which is located at $\epsilon=0$ in the
case $\xi=0$. Finally, when $|eV|\ll\Gamma$, there is photon absorption with
weak excitations of the quantum dot, that is with excitation energies
$\Delta\epsilon\lesssim\Gamma$, and involves tunneling processes after which
the quasiparticle energy in the final state is increased by
$\Delta\epsilon_{qp}\lesssim\Gamma$. The first type of such tunneling includes
\begin{figure}
\includegraphics[width=8.0 cm]{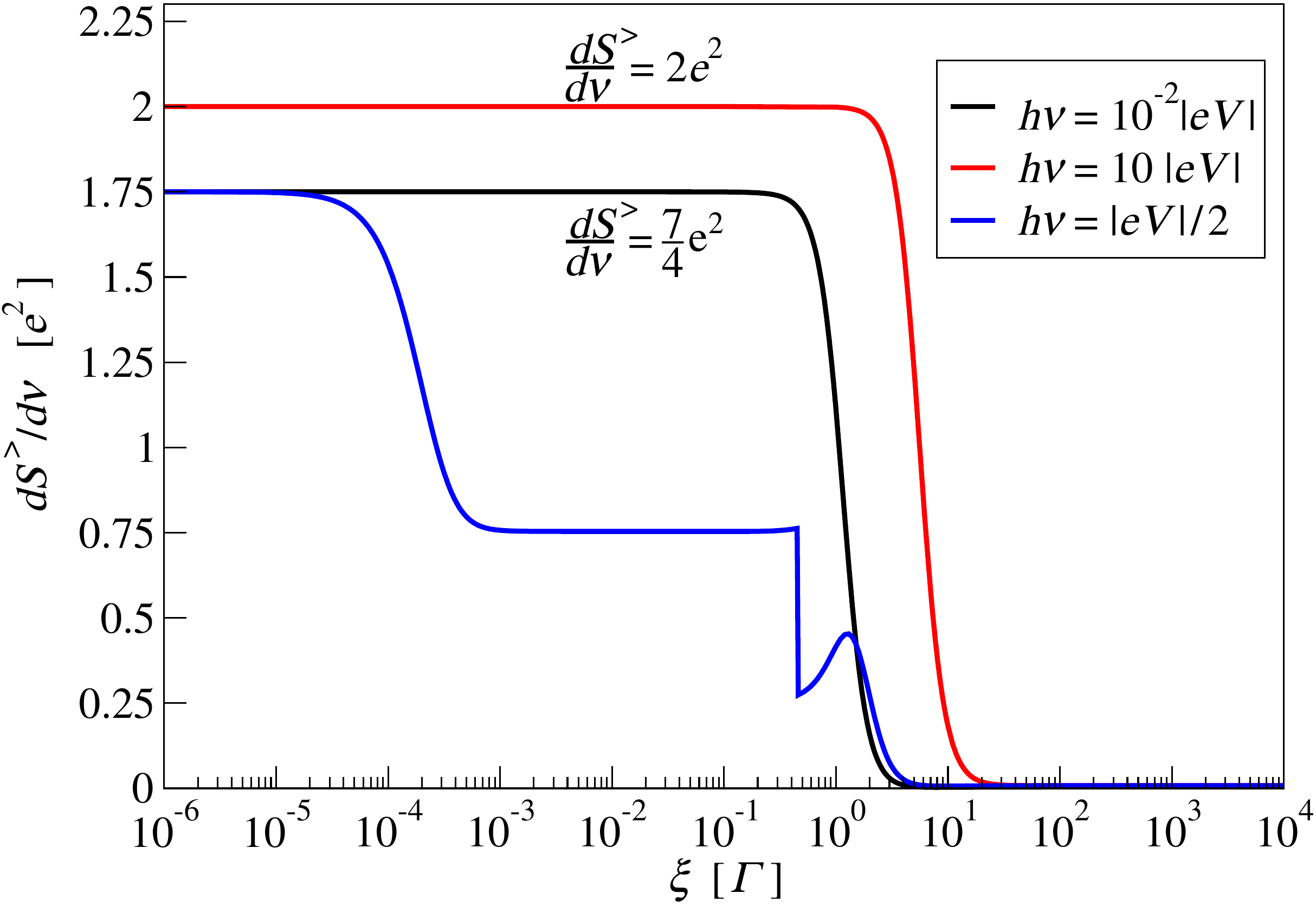}
\caption{\label{figure_5} The differential quantum noise
  $\partial S^>(\nu,V)/\partial\nu$ as a function of the overlap energy $\xi$
  is shown for $|eV|/\Gamma=10^{-4}$. Here, as in Fig. \ref{figure_2},
  $k_\text{B}T/\Gamma=10^{-12}$, $\epsilon_d/\Gamma=8.0$,
  $|\eta|/\Gamma=10^3$. The black curve corresponds to $h\nu/\Gamma=10^{-6}$,
  {\it i.e.} to the case $h\nu<|eV|$. The red curve corresponds to
  $h\nu/\Gamma=10^{-3}$, {\it i.e.} to the case $\Gamma>h\nu>|eV|$. The blue
  curve corresponds to $h\nu/\Gamma=5\cdot 10^{-5}$, {\it i.e.} to the case
  $h\nu=|eV|/2$ and, therefore, describes the behavior of the antiresonance
  minimum.}
\end{figure}
tunneling processes {\it left contact} $\rightarrow$ {\it quantum dot}
$\rightarrow$ {\it left contact} which are involved in photon absorption at
frequencies $h\nu<|eV|$ leading to the formation of the first low-energy
plateau, $\partial S^>(\nu,V)/\partial\nu=7e^2/4$, shown in
Fig. \ref{figure_2}. The second type of such tunneling includes tunneling
processes {\it left contact} $\rightarrow$ {\it quantum dot} $\rightarrow$
{\it right contact} which are involved in photon absorption at frequencies
$h\nu>|eV|$. Tunneling processes of the second type add to the tunneling
processes of the first type, which are present also for $h\nu>|eV|$, and lead
to the increase of the differential noise up to the second low-energy plateau,
$\partial S^>(\nu,V)/\partial\nu=2e^2$, shown in Fig. \ref{figure_2}. As we
have seen in Fig. \ref{figure_2}, indeed, at $h\nu=|eV|$ there is a jump in
$\partial S^>(\nu,V)/\partial\nu$. Why the two low-energy plateaus are equal
to exactly $7e^2/4$ and $2e^2$ is, however, an open question representing a
challenge for future research elaborating more analytical approaches.
\begin{figure}
\includegraphics[width=8.0 cm]{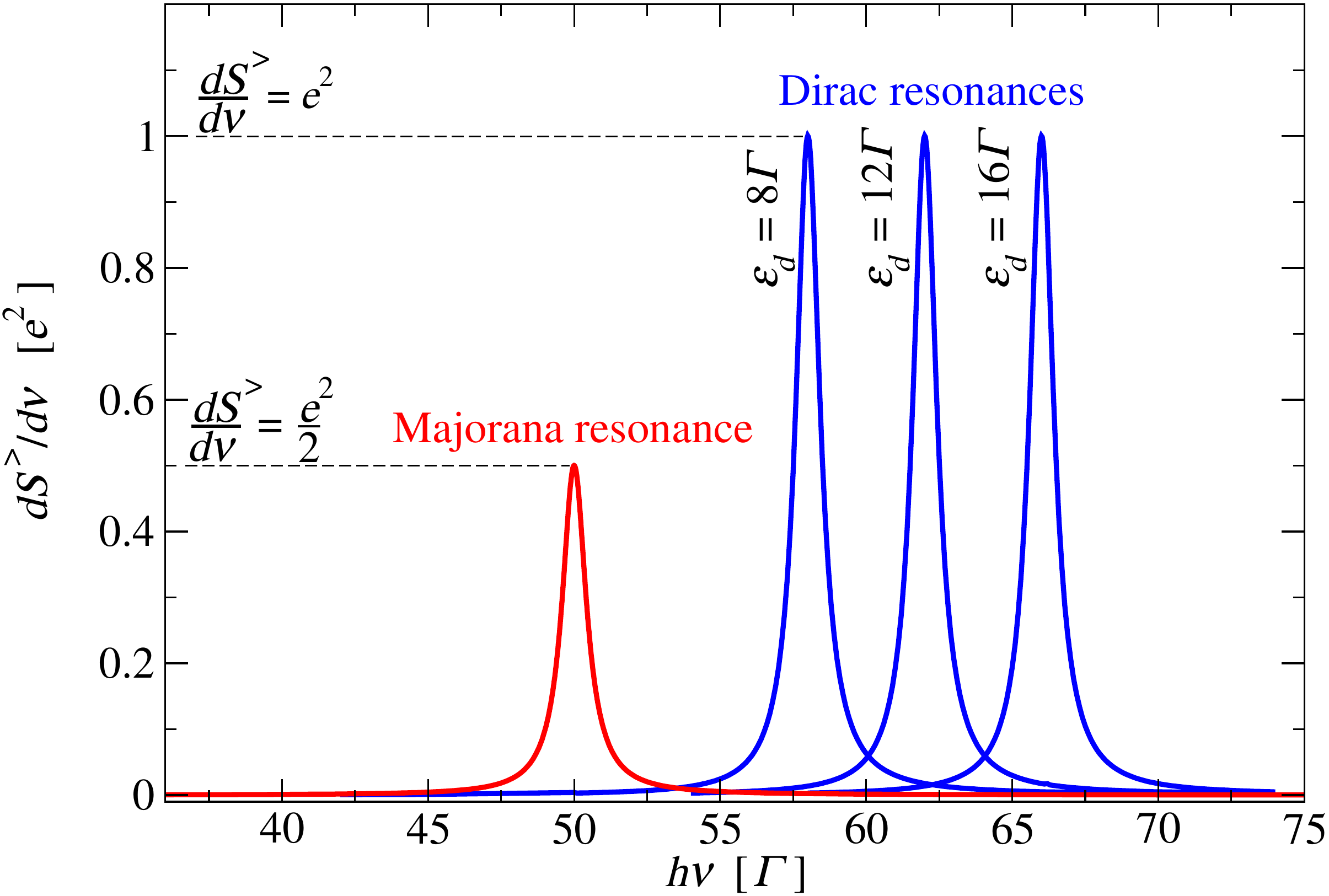}
\caption{\label{figure_6} The differential quantum noise
  $\partial S^>(\nu,V)/\partial\nu$ as a function of the frequency $\nu$ is
  shown for $|eV|/\Gamma=100.0$. Here, as in Fig. \ref{figure_2},
  $k_\text{B}T/\Gamma=10^{-12}$, $|\eta|/\Gamma=10^3$. The red curve is for
  the case of the well separated Majorana modes, $\xi/\Gamma=10^{-4}$. In this
  case the Majorana resonance at $h\nu=|eV|/2$ is universal, {\it i.e.} it is
  independent of $\epsilon_d$ controlled by a gate voltage. The blue curves
  are for the case of strongly, $\xi/\Gamma=10^6$, overlapping Majorana modes
  merging into a single Dirac fermion. In this case the Dirac resonance is not
  universal. Its location depends on $\epsilon_d$ as $\epsilon_d+|eV|/2$. The
  three Dirac resonances are shown for $\epsilon_d/\Gamma=8,12,16$ from left
  to right, respectively.}
\end{figure}

It is important to look at what will happen with the above discussed universal
Majorana features of finite frequency noise if the Majorana bound states are
absent. To this end we consider the scenario when the two Majorana bound
states disappear via their significant overlap resulting in appearance of a
single Dirac fermion instead of the unpaired Majoranas. Physically this is
relevant for short topological superconductors. This situation can be modeled
via increasing the overlap energy $\xi$. Different universal Majorana features
shown above disappear in different ways which we demonstrate below.

As shown in Fig. \ref{figure_5}, in the case $|eV|\ll\Gamma$ the two
low-frequency plateaus (see Fig. \ref{figure_2}) with the universal values
$\partial S^>(\nu,V)/\partial\nu=7e^2/4$ (black curve) for $h\nu<|eV|$ and
$\partial S^>(\nu,V)/\partial\nu=2e^2$ (red curve) for $h\nu>|eV|$ are totally
ruined when the two Majorana bound states merge into a single Dirac
fermion. The blue curve shows what happens with the antiresonance (see
Fig. \ref{figure_2}) located at $h\nu=|eV|/2$. As one can see, the
antiresonance is also destroyed when the Majorana bound states strongly
overlap. It is interesting to note that for the extremely well,
$\xi/\Gamma<10^{-5}$, separated Majorana bound states the antiresonance fully
disappears. But for a stronger overlap, such that all other universal Majorana
features discussed above do not change, it can already be observed as is the
case in Fig. \ref{figure_2} where $\xi/\Gamma=10^{-4}$.
\begin{figure}
\includegraphics[width=8.0 cm]{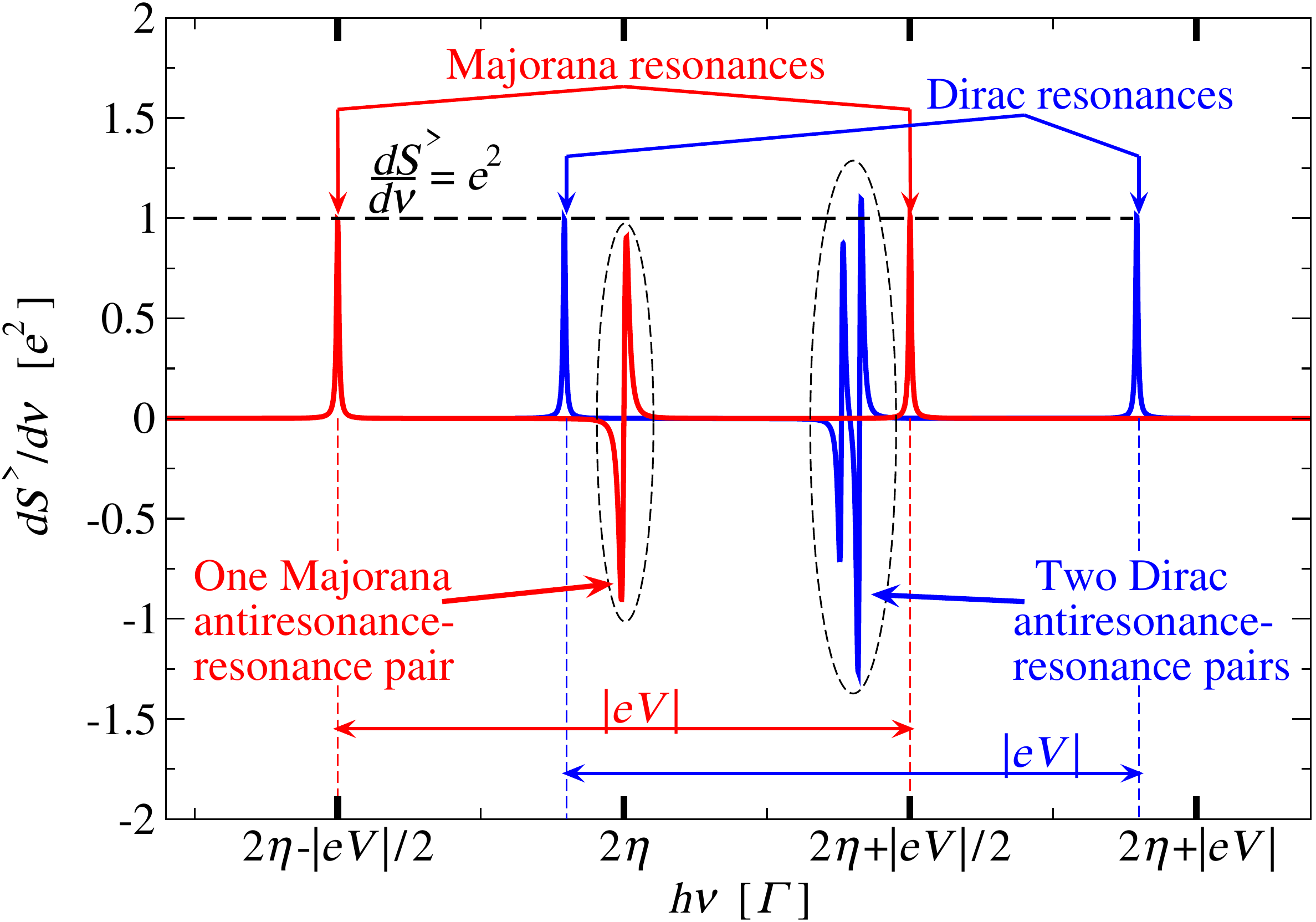}
\caption{\label{figure_7} The differential quantum noise
  $\partial S^>(\nu,V)/\partial\nu$ as a function of the frequency $\nu$ is
  shown for $|eV|/\Gamma=100.0$. Here, as in Fig. \ref{figure_2},
  $k_\text{B}T/\Gamma=10^{-12}$, $|\eta|/\Gamma=10^3$. The red curve is for
  the case of the well separated Majorana modes, $\xi/\Gamma=10^{-4}$. The
  blue curve is for the case of significantly, $\xi/\Gamma=400$, overlapping
  Majorana modes merging into a single Dirac fermion.}
\end{figure}

In Fig. \ref{figure_6} we demonstrate what happens with the Majorana resonance
$\partial S^>(\nu,V)/\partial\nu=e^2/2$ located at $h\nu=|eV|/2$ for
$|eV|\gg\Gamma$. When the two Majorana bound states strongly overlap and form
a single Dirac fermion, instead of the Majorana resonance there develops a
Dirac resonance which is twice stronger,
$\partial S^>(\nu,V)/\partial\nu=e^2$. While the Majorana resonance is
universal, that is independent of $\epsilon_d$, the Dirac resonance is not
universal because its location is specified by $\epsilon_d$ as
$\epsilon_d+|eV|/2$. Therefore, if in an experiment one continuously changes
the gate voltage controlling $\epsilon_d$, this Dirac resonance will
continuously flow according to that change as shown in Fig. \ref{figure_6} for
three different values of $\epsilon_d$.

Finally, Fig. \ref{figure_7} illustrates for $|eV|\gg\Gamma$ how the two
universal ({\it i.e.} independent of $\epsilon_d$) Majorana resonances (red
curve) $\partial S^>(\nu,V)/\partial\nu=e^2$ located at
$h\nu=2|\eta|\mp|eV|/2$ turn into two Dirac resonances (blue curve). These
Dirac resonances have the same magnitude $e^2$ as the Majorana resonances but
they are shifted from the points $2|\eta|\mp|eV|/2$. This shift depends on
$\xi$ but the distance between the Dirac resonances remains equal to
$|eV|$. For larger values of $\xi$ the Dirac resonances will shift to higher
frequencies, $h\nu\gg 2|\eta|\mp|eV|/2$. Moreover, as we have seen in
Fig. \ref{figure_4}, there is one (odd number) universal Majorana
antiresonance-resonance pair centered at $2|\eta|$ as shown in
Fig. \ref{figure_7} by the dashed ellipse around the red curve. However, when
the two Majorana bound states significantly overlap and form a single Dirac
fermion, there develop two (even number) Dirac antiresonance-resonance pairs
located just in the center between the two Dirac resonances as it is shown in
Fig. \ref{figure_7} by the dashed ellipse around the blue curve.
\section{Conclusion}\label{concl}
In conclusion, we have explored the photon absorption spectra governed by Majorana
zero modes in a mesoscopic setup and predicted their universal nature revealed
via our extensive numerical computations of the differential quantum noise
over a large range of frequencies and bias voltages. Although finite frequency
noise in specific settings, {\it e.g.}, in a topological superconductor-normal
metal structure \cite{Valentini_2016,Bathellier_2019}, may be used as a tool
to distinguish between Majorana zero modes and zero-energy ordinary Andreev
bound states, universal fingerprints of the Majorana finite frequency quantum
noise have never been discussed. At the same time universality of finite
frequency quantum noise is of particular relevance for experiments on Majorana
zero modes and their applications. Our setup is especially attractive for
state-of-the-art experiments involving quantum noise detectors able to reach
the quantum limit and separately access photon absorption and emission
spectra. One reason is that single-electron transistors and noninteracting
resonant-level quantum noise detectors, where only one energy level
contributes to transport (and thus Coulomb-blockade effects may be neglected),
have already been proposed
\cite{Averin_2000,Clerk_2004,Mozyrsky_2004}. Importantly, these mesoscopic
quantum noise detectors are able to reach the quantum limit in nonequilibrium
states induced by both small and large bias voltages. Another reason is that
our results for quantum dot setups, where only one energy level contributes to
transport, should be robust against disorder. Indeed, disorder might affect
the Majorana transport through the quantum dot in two ways. First, the
topological superconductor itself can be disordered and interplay between
topology and disorder specifies when the Majorana bound states are present or
absent at the ends of the topological superconductor. In this case the finite
frequency noise will follow that interplay between topology and disorder and
our results presented here will show either universal Majorana or nonuniversal
Dirac behavior depending on whether the Majorana bound states are,
respectively, present or absent at the ends of the topological
superconductor. Here, for a typical setup, we neglect a feedback of a small
quantum dot on a large topological superconductor. Otherwise, if such a
feedback becomes significant, there will appear a complicated interplay
between topology and disorder of the topological superconductor and
nonequilibrium states of the quantum dot. This complicated nonequilibrium
dynamics involving topology and disorder would be a challenge for future
research. Second, there can be a particular realization of disorder in the
quantum dot whose discrete energy spectrum is specified by that particular
disorder. In this case one may use the gate voltage to tune the chemical
potential around a single-particle energy level $\epsilon_d$ to reach the
transport regime where only one energy level $\epsilon_d$ essentially
contributes to transport as discussed above. Then our results on finite
frequency noise will be applicable again. This assumes that for a particular
realization of disorder the distances between the energy levels of the quantum
dot should be large enough because otherwise more than one energy level will
contribute to transport. Such a multilevel regime would be another challenge
for future research.

Both of the above discussed reasons imply that the results presented in this
work pertain to realistic systems and thus they are of immediate interest for
experiments on Majorana finite frequency noise in the quantum limit.

\section*{Acknowledgments}
The author thanks Milena Grifoni, Andreas K. H\"uttel and Wataru Izumida for
invaluable discussions.


\begin{thebibliography}{31}%
\makeatletter
\providecommand \@ifxundefined [1]{%
 \@ifx{#1\undefined}
}%
\providecommand \@ifnum [1]{%
 \ifnum #1\expandafter \@firstoftwo
 \else \expandafter \@secondoftwo
 \fi
}%
\providecommand \@ifx [1]{%
 \ifx #1\expandafter \@firstoftwo
 \else \expandafter \@secondoftwo
 \fi
}%
\providecommand \natexlab [1]{#1}%
\providecommand \enquote  [1]{``#1''}%
\providecommand \bibnamefont  [1]{#1}%
\providecommand \bibfnamefont [1]{#1}%
\providecommand \citenamefont [1]{#1}%
\providecommand \href@noop [0]{\@secondoftwo}%
\providecommand \href [0]{\begingroup \@sanitize@url \@href}%
\providecommand \@href[1]{\@@startlink{#1}\@@href}%
\providecommand \@@href[1]{\endgroup#1\@@endlink}%
\providecommand \@sanitize@url [0]{\catcode `\\12\catcode `\$12\catcode
  `\&12\catcode `\#12\catcode `\^12\catcode `\_12\catcode `\%12\relax}%
\providecommand \@@startlink[1]{}%
\providecommand \@@endlink[0]{}%
\providecommand \url  [0]{\begingroup\@sanitize@url \@url }%
\providecommand \@url [1]{\endgroup\@href {#1}{\urlprefix }}%
\providecommand \urlprefix  [0]{URL }%
\providecommand \Eprint [0]{\href }%
\providecommand \doibase [0]{http://dx.doi.org/}%
\providecommand \selectlanguage [0]{\@gobble}%
\providecommand \bibinfo  [0]{\@secondoftwo}%
\providecommand \bibfield  [0]{\@secondoftwo}%
\providecommand \translation [1]{[#1]}%
\providecommand \BibitemOpen [0]{}%
\providecommand \bibitemStop [0]{}%
\providecommand \bibitemNoStop [0]{.\EOS\space}%
\providecommand \EOS [0]{\spacefactor3000\relax}%
\providecommand \BibitemShut  [1]{\csname bibitem#1\endcsname}%
\let\auto@bib@innerbib\@empty
\bibitem [{\citenamefont {\text{Yu.} Kitaev}(2001)}]{Kitaev_2001}%
  \BibitemOpen
  \bibfield  {author} {\bibinfo {author} {\bibfnamefont {A.}~\bibnamefont
  {\text{Yu.} Kitaev}},\ }\bibfield  {title} {\enquote {\bibinfo {title}
  {Unpaired {M}ajorana fermions in quantum wires},}\ }\href@noop {} {\bibfield
  {journal} {\bibinfo  {journal} {Phys.-Usp.}\ }\textbf {\bibinfo {volume}
  {44}},\ \bibinfo {pages} {131} (\bibinfo {year} {2001})}\BibitemShut
  {NoStop}%
\bibitem [{\citenamefont {Majorana}(1937)}]{Majorana_1937}%
  \BibitemOpen
  \bibfield  {author} {\bibinfo {author} {\bibfnamefont {E.}~\bibnamefont
  {Majorana}},\ }\bibfield  {title} {\enquote {\bibinfo {title} {Teoria
  simmetrica dell'elettrone e del positrone},}\ }\href@noop {} {\bibfield
  {journal} {\bibinfo  {journal} {Nuovo Cimento}\ }\textbf {\bibinfo {volume}
  {14}},\ \bibinfo {pages} {171} (\bibinfo {year} {1937})}\BibitemShut
  {NoStop}%
\bibitem [{\citenamefont {Alicea}(2012)}]{Alicea_2012}%
  \BibitemOpen
  \bibfield  {author} {\bibinfo {author} {\bibfnamefont {J.}~\bibnamefont
  {Alicea}},\ }\bibfield  {title} {\enquote {\bibinfo {title} {New directions
  in the pursuit of {M}ajorana fermions in solid state systems},}\ }\href@noop
  {} {\bibfield  {journal} {\bibinfo  {journal} {Rep. Prog. Phys.}\ }\textbf
  {\bibinfo {volume} {75}},\ \bibinfo {pages} {076501} (\bibinfo {year}
  {2012})}\BibitemShut {NoStop}%
\bibitem [{\citenamefont {Leijnse}\ and\ \citenamefont
  {Flensberg}(2012)}]{Flensberg_2012}%
  \BibitemOpen
  \bibfield  {author} {\bibinfo {author} {\bibfnamefont {M.}~\bibnamefont
  {Leijnse}}\ and\ \bibinfo {author} {\bibfnamefont {K.}~\bibnamefont
  {Flensberg}},\ }\bibfield  {title} {\enquote {\bibinfo {title} {Introduction
  to topological superconductivity and {M}ajorana fermions},}\ }\href@noop {}
  {\bibfield  {journal} {\bibinfo  {journal} {Semicond. Sci. Technol.}\
  }\textbf {\bibinfo {volume} {27}},\ \bibinfo {pages} {124003} (\bibinfo
  {year} {2012})}\BibitemShut {NoStop}%
\bibitem [{\citenamefont {Sato}\ and\ \citenamefont
  {Fujimoto}(2016)}]{Sato_2016}%
  \BibitemOpen
  \bibfield  {author} {\bibinfo {author} {\bibfnamefont {M.}~\bibnamefont
  {Sato}}\ and\ \bibinfo {author} {\bibfnamefont {S.}~\bibnamefont
  {Fujimoto}},\ }\bibfield  {title} {\enquote {\bibinfo {title} {Majorana
  fermions and topology in superconductors},}\ }\href@noop {} {\bibfield
  {journal} {\bibinfo  {journal} {J. Phys. Soc. Japan}\ }\textbf {\bibinfo
  {volume} {85}},\ \bibinfo {pages} {072001} (\bibinfo {year}
  {2016})}\BibitemShut {NoStop}%
\bibitem [{\citenamefont {Lutchyn}\ \emph {et~al.}(2018)\citenamefont
  {Lutchyn}, \citenamefont {Bakkers}, \citenamefont {Kouwenhoven},
  \citenamefont {Krogstrup}, \citenamefont {Marcus},\ and\ \citenamefont
  {Oreg}}]{Lutchyn_2018}%
  \BibitemOpen
  \bibfield  {author} {\bibinfo {author} {\bibfnamefont {R.~M.}\ \bibnamefont
  {Lutchyn}}, \bibinfo {author} {\bibfnamefont {E.~P. A.~M.}\ \bibnamefont
  {Bakkers}}, \bibinfo {author} {\bibfnamefont {L.~P.}\ \bibnamefont
  {Kouwenhoven}}, \bibinfo {author} {\bibfnamefont {P.}~\bibnamefont
  {Krogstrup}}, \bibinfo {author} {\bibfnamefont {C.~M.}\ \bibnamefont
  {Marcus}}, \ and\ \bibinfo {author} {\bibfnamefont {Y.}~\bibnamefont
  {Oreg}},\ }\bibfield  {title} {\enquote {\bibinfo {title} {Majorana zero
  modes in superconductor-semiconductor heterostructures},}\ }\href@noop {}
  {\bibfield  {journal} {\bibinfo  {journal} {Nat. Rev. Mater.}\ }\textbf
  {\bibinfo {volume} {3}},\ \bibinfo {pages} {52} (\bibinfo {year}
  {2018})}\BibitemShut {NoStop}%
\bibitem [{\citenamefont {Mourik}\ \emph {et~al.}(2012)\citenamefont {Mourik},
  \citenamefont {Zuo}, \citenamefont {Frolov}, \citenamefont {Plissard},
  \citenamefont {Bakkers},\ and\ \citenamefont {Kouwenhoven}}]{Mourik_2012}%
  \BibitemOpen
  \bibfield  {author} {\bibinfo {author} {\bibfnamefont {V.}~\bibnamefont
  {Mourik}}, \bibinfo {author} {\bibfnamefont {K.}~\bibnamefont {Zuo}},
  \bibinfo {author} {\bibfnamefont {S.~M.}\ \bibnamefont {Frolov}}, \bibinfo
  {author} {\bibfnamefont {S.~R.}\ \bibnamefont {Plissard}}, \bibinfo {author}
  {\bibfnamefont {E.~P. A.~M.}\ \bibnamefont {Bakkers}}, \ and\ \bibinfo
  {author} {\bibfnamefont {L.~P.}\ \bibnamefont {Kouwenhoven}},\ }\bibfield
  {title} {\enquote {\bibinfo {title} {Signatures of {M}ajorana fermions in
  hybrid superconductor-semiconductor nanowire devices},}\ }\href@noop {}
  {\bibfield  {journal} {\bibinfo  {journal} {Science}\ }\textbf {\bibinfo
  {volume} {336}},\ \bibinfo {pages} {1003} (\bibinfo {year}
  {2012})}\BibitemShut {NoStop}%
\bibitem [{\citenamefont {Albrecht}\ \emph {et~al.}(2016)\citenamefont
  {Albrecht}, \citenamefont {Higginbotham}, \citenamefont {Madsen},
  \citenamefont {Kuemmeth}, \citenamefont {Jespersen}, \citenamefont
  {Nyg{\r{a}}rd}, \citenamefont {Krogstrup},\ and\ \citenamefont
  {Marcus}}]{Albrecht_2016}%
  \BibitemOpen
  \bibfield  {author} {\bibinfo {author} {\bibfnamefont {S.~M.}\ \bibnamefont
  {Albrecht}}, \bibinfo {author} {\bibfnamefont {A.~P.}\ \bibnamefont
  {Higginbotham}}, \bibinfo {author} {\bibfnamefont {M.}~\bibnamefont
  {Madsen}}, \bibinfo {author} {\bibfnamefont {F.}~\bibnamefont {Kuemmeth}},
  \bibinfo {author} {\bibfnamefont {T.~S.}\ \bibnamefont {Jespersen}}, \bibinfo
  {author} {\bibfnamefont {J.}~\bibnamefont {Nyg{\r{a}}rd}}, \bibinfo {author}
  {\bibfnamefont {P.}~\bibnamefont {Krogstrup}}, \ and\ \bibinfo {author}
  {\bibfnamefont {C.~M.}\ \bibnamefont {Marcus}},\ }\bibfield  {title}
  {\enquote {\bibinfo {title} {Exponential protection of zero modes in
  {M}ajorana islands},}\ }\href@noop {} {\bibfield  {journal} {\bibinfo
  {journal} {Nature}\ }\textbf {\bibinfo {volume} {531}},\ \bibinfo {pages}
  {206} (\bibinfo {year} {2016})}\BibitemShut {NoStop}%
\bibitem [{\citenamefont {Zhang}\ \emph {et~al.}(2018)\citenamefont {Zhang},
  \citenamefont {Liu}, \citenamefont {Gazibegovic}, \citenamefont {Xu},
  \citenamefont {Logan}, \citenamefont {Wang}, \citenamefont {van Loo},
  \citenamefont {Bommer}, \citenamefont {de~Moor}, \citenamefont {Car},
  \citenamefont {het Veld}, \citenamefont {van Veldhoven}, \citenamefont
  {Koelling}, \citenamefont {Verheijen}, \citenamefont {Pendharkar},
  \citenamefont {Pennachio}, \citenamefont {Shojaei}, \citenamefont {Lee},
  \citenamefont {Palmstr{\o}m}, \citenamefont {Bakkers}, \citenamefont
  {Sarma},\ and\ \citenamefont {Kouwenhoven}}]{Zhang_2018}%
  \BibitemOpen
  \bibfield  {author} {\bibinfo {author} {\bibfnamefont {H.}~\bibnamefont
  {Zhang}}, \bibinfo {author} {\bibfnamefont {C.-X.}\ \bibnamefont {Liu}},
  \bibinfo {author} {\bibfnamefont {S.}~\bibnamefont {Gazibegovic}}, \bibinfo
  {author} {\bibfnamefont {D.}~\bibnamefont {Xu}}, \bibinfo {author}
  {\bibfnamefont {J.~A.}\ \bibnamefont {Logan}}, \bibinfo {author}
  {\bibfnamefont {G.}~\bibnamefont {Wang}}, \bibinfo {author} {\bibfnamefont
  {N.}~\bibnamefont {van Loo}}, \bibinfo {author} {\bibfnamefont {J.~D.~S.}\
  \bibnamefont {Bommer}}, \bibinfo {author} {\bibfnamefont {M.~W.~A.}\
  \bibnamefont {de~Moor}}, \bibinfo {author} {\bibfnamefont {D.}~\bibnamefont
  {Car}}, \bibinfo {author} {\bibfnamefont {R.~L. M.~O.}\ \bibnamefont {het
  Veld}}, \bibinfo {author} {\bibfnamefont {P.~J.}\ \bibnamefont {van
  Veldhoven}}, \bibinfo {author} {\bibfnamefont {S.}~\bibnamefont {Koelling}},
  \bibinfo {author} {\bibfnamefont {M.~A.}\ \bibnamefont {Verheijen}}, \bibinfo
  {author} {\bibfnamefont {M.}~\bibnamefont {Pendharkar}}, \bibinfo {author}
  {\bibfnamefont {D.~J.}\ \bibnamefont {Pennachio}}, \bibinfo {author}
  {\bibfnamefont {B.}~\bibnamefont {Shojaei}}, \bibinfo {author} {\bibfnamefont
  {J.~S.}\ \bibnamefont {Lee}}, \bibinfo {author} {\bibfnamefont {C.~J.}\
  \bibnamefont {Palmstr{\o}m}}, \bibinfo {author} {\bibfnamefont {E.~P. A.~M.}\
  \bibnamefont {Bakkers}}, \bibinfo {author} {\bibfnamefont {S.~D.}\
  \bibnamefont {Sarma}}, \ and\ \bibinfo {author} {\bibfnamefont {L.~P.}\
  \bibnamefont {Kouwenhoven}},\ }\bibfield  {title} {\enquote {\bibinfo {title}
  {Quantized {M}ajorana conductance},}\ }\href@noop {} {\bibfield  {journal}
  {\bibinfo  {journal} {Nature}\ }\textbf {\bibinfo {volume} {556}},\ \bibinfo
  {pages} {74} (\bibinfo {year} {2018})}\BibitemShut {NoStop}%
\bibitem [{\citenamefont {Smirnov}(2015)}]{Smirnov_2015}%
  \BibitemOpen
  \bibfield  {author} {\bibinfo {author} {\bibfnamefont {S.}~\bibnamefont
  {Smirnov}},\ }\bibfield  {title} {\enquote {\bibinfo {title} {Majorana
  tunneling entropy},}\ }\href@noop {} {\bibfield  {journal} {\bibinfo
  {journal} {Phys.\ Rev.\ B}\ }\textbf {\bibinfo {volume} {92}},\ \bibinfo
  {pages} {195312} (\bibinfo {year} {2015})}\BibitemShut {NoStop}%
\bibitem [{\citenamefont {Hartman}\ \emph {et~al.}(2018)\citenamefont
  {Hartman}, \citenamefont {Olsen}, \citenamefont {L{\"u}scher}, \citenamefont
  {Samani}, \citenamefont {Fallahi}, \citenamefont {Gardner}, \citenamefont
  {Manfra},\ and\ \citenamefont {Folk}}]{Hartman_2018}%
  \BibitemOpen
  \bibfield  {author} {\bibinfo {author} {\bibfnamefont {N.}~\bibnamefont
  {Hartman}}, \bibinfo {author} {\bibfnamefont {C.}~\bibnamefont {Olsen}},
  \bibinfo {author} {\bibfnamefont {S.}~\bibnamefont {L{\"u}scher}}, \bibinfo
  {author} {\bibfnamefont {M.}~\bibnamefont {Samani}}, \bibinfo {author}
  {\bibfnamefont {S.}~\bibnamefont {Fallahi}}, \bibinfo {author} {\bibfnamefont
  {G.~C.}\ \bibnamefont {Gardner}}, \bibinfo {author} {\bibfnamefont
  {M.}~\bibnamefont {Manfra}}, \ and\ \bibinfo {author} {\bibfnamefont
  {J.}~\bibnamefont {Folk}},\ }\bibfield  {title} {\enquote {\bibinfo {title}
  {Direct entropy measurement in a mesoscopic quantum system},}\ }\href@noop {}
  {\bibfield  {journal} {\bibinfo  {journal} {Nature Physics}\ }\textbf
  {\bibinfo {volume} {14}},\ \bibinfo {pages} {1083} (\bibinfo {year}
  {2018})}\BibitemShut {NoStop}%
\bibitem [{\citenamefont {\text{Yu.} Kitaev}(2003)}]{Kitaev_2003}%
  \BibitemOpen
  \bibfield  {author} {\bibinfo {author} {\bibfnamefont {A.}~\bibnamefont
  {\text{Yu.} Kitaev}},\ }\bibfield  {title} {\enquote {\bibinfo {title}
  {Fault-tolerant quantum computation by anyons},}\ }\href@noop {} {\bibfield
  {journal} {\bibinfo  {journal} {Ann. Phys.}\ }\textbf {\bibinfo {volume}
  {303}},\ \bibinfo {pages} {2} (\bibinfo {year} {2003})}\BibitemShut {NoStop}%
\bibitem [{\citenamefont {Nayak}\ \emph {et~al.}(2008)\citenamefont {Nayak},
  \citenamefont {Simon}, \citenamefont {Stern}, \citenamefont {Freedman},\ and\
  \citenamefont {\text{Das Sarma}}}]{Nayak_2008}%
  \BibitemOpen
  \bibfield  {author} {\bibinfo {author} {\bibfnamefont {C.}~\bibnamefont
  {Nayak}}, \bibinfo {author} {\bibfnamefont {S.~H.}\ \bibnamefont {Simon}},
  \bibinfo {author} {\bibfnamefont {A.}~\bibnamefont {Stern}}, \bibinfo
  {author} {\bibfnamefont {M.}~\bibnamefont {Freedman}}, \ and\ \bibinfo
  {author} {\bibfnamefont {S.}~\bibnamefont {\text{Das Sarma}}},\ }\bibfield
  {title} {\enquote {\bibinfo {title} {Non-abelian anyons and topological
  quantum computation},}\ }\href@noop {} {\bibfield  {journal} {\bibinfo
  {journal} {Rev.\ Mod.\ Phys.}\ }\textbf {\bibinfo {volume} {80}},\ \bibinfo
  {pages} {1083} (\bibinfo {year} {2008})}\BibitemShut {NoStop}%
\bibitem [{\citenamefont {Alicea}\ \emph {et~al.}(2011)\citenamefont {Alicea},
  \citenamefont {Oreg}, \citenamefont {Refael}, \citenamefont {von Oppen},\
  and\ \citenamefont {Fisher}}]{Alicea_2011}%
  \BibitemOpen
  \bibfield  {author} {\bibinfo {author} {\bibfnamefont {J.}~\bibnamefont
  {Alicea}}, \bibinfo {author} {\bibfnamefont {Y.}~\bibnamefont {Oreg}},
  \bibinfo {author} {\bibfnamefont {G.}~\bibnamefont {Refael}}, \bibinfo
  {author} {\bibfnamefont {F.}~\bibnamefont {von Oppen}}, \ and\ \bibinfo
  {author} {\bibfnamefont {M.~P.~A.}\ \bibnamefont {Fisher}},\ }\bibfield
  {title} {\enquote {\bibinfo {title} {Non-{A}belian statistics and topological
  quantum information processing in 1{D} wire networks},}\ }\href@noop {}
  {\bibfield  {journal} {\bibinfo  {journal} {Nature Physics}\ }\textbf
  {\bibinfo {volume} {7}},\ \bibinfo {pages} {412} (\bibinfo {year}
  {2011})}\BibitemShut {NoStop}%
\bibitem [{\citenamefont {Aasen}\ \emph {et~al.}(2016)\citenamefont {Aasen},
  \citenamefont {Hell}, \citenamefont {Mishmash}, \citenamefont {Higginbotham},
  \citenamefont {Danon}, \citenamefont {Leijnse}, \citenamefont {Jespersen},
  \citenamefont {Folk}, \citenamefont {Marcus}, \citenamefont {Flensberg},\
  and\ \citenamefont {Alicea}}]{Aasen_2016}%
  \BibitemOpen
  \bibfield  {author} {\bibinfo {author} {\bibfnamefont {D.}~\bibnamefont
  {Aasen}}, \bibinfo {author} {\bibfnamefont {M.}~\bibnamefont {Hell}},
  \bibinfo {author} {\bibfnamefont {R.~V.}\ \bibnamefont {Mishmash}}, \bibinfo
  {author} {\bibfnamefont {A.}~\bibnamefont {Higginbotham}}, \bibinfo {author}
  {\bibfnamefont {J.}~\bibnamefont {Danon}}, \bibinfo {author} {\bibfnamefont
  {M.}~\bibnamefont {Leijnse}}, \bibinfo {author} {\bibfnamefont {T.~S.}\
  \bibnamefont {Jespersen}}, \bibinfo {author} {\bibfnamefont {J.~A.}\
  \bibnamefont {Folk}}, \bibinfo {author} {\bibfnamefont {C.~M.}\ \bibnamefont
  {Marcus}}, \bibinfo {author} {\bibfnamefont {K.}~\bibnamefont {Flensberg}}, \
  and\ \bibinfo {author} {\bibfnamefont {J.}~\bibnamefont {Alicea}},\
  }\bibfield  {title} {\enquote {\bibinfo {title} {Milestones toward
  {M}ajorana-based quantum computing},}\ }\href@noop {} {\bibfield  {journal}
  {\bibinfo  {journal} {Phys.\ Rev.\ X}\ }\textbf {\bibinfo {volume} {6}},\
  \bibinfo {pages} {031016} (\bibinfo {year} {2016})}\BibitemShut {NoStop}%
\bibitem [{\citenamefont {Vijay}\ and\ \citenamefont {Fu}(2016)}]{Vijay_2016}%
  \BibitemOpen
  \bibfield  {author} {\bibinfo {author} {\bibfnamefont {S.}~\bibnamefont
  {Vijay}}\ and\ \bibinfo {author} {\bibfnamefont {L.}~\bibnamefont {Fu}},\
  }\bibfield  {title} {\enquote {\bibinfo {title} {Teleportation-based quantum
  information processing with {M}ajorana zero modes},}\ }\href@noop {}
  {\bibfield  {journal} {\bibinfo  {journal} {Phys.\ Rev.\ B}\ }\textbf
  {\bibinfo {volume} {94}},\ \bibinfo {pages} {235446} (\bibinfo {year}
  {2016})}\BibitemShut {NoStop}%
\bibitem [{\citenamefont {Pedrocchi}\ and\ \citenamefont
  {DiVincenzo}(2015)}]{Pedrocchi_2015}%
  \BibitemOpen
  \bibfield  {author} {\bibinfo {author} {\bibfnamefont {F.~L.}\ \bibnamefont
  {Pedrocchi}}\ and\ \bibinfo {author} {\bibfnamefont {D.~P.}\ \bibnamefont
  {DiVincenzo}},\ }\bibfield  {title} {\enquote {\bibinfo {title} {Majorana
  braiding with thermal noise},}\ }\href@noop {} {\bibfield  {journal}
  {\bibinfo  {journal} {Phys.\ Rev.\ Lett.}\ }\textbf {\bibinfo {volume}
  {115}},\ \bibinfo {pages} {120402} (\bibinfo {year} {2015})}\BibitemShut
  {NoStop}%
\bibitem [{\citenamefont {Gharavi}\ \emph {et~al.}(2016)\citenamefont
  {Gharavi}, \citenamefont {Hoving},\ and\ \citenamefont
  {Baugh}}]{Gharavi_2016}%
  \BibitemOpen
  \bibfield  {author} {\bibinfo {author} {\bibfnamefont {K.}~\bibnamefont
  {Gharavi}}, \bibinfo {author} {\bibfnamefont {D.}~\bibnamefont {Hoving}}, \
  and\ \bibinfo {author} {\bibfnamefont {J.}~\bibnamefont {Baugh}},\ }\bibfield
   {title} {\enquote {\bibinfo {title} {Readout of {M}ajorana parity states
  using a quantum dot},}\ }\href@noop {} {\bibfield  {journal} {\bibinfo
  {journal} {Phys.\ Rev.\ B}\ }\textbf {\bibinfo {volume} {94}},\ \bibinfo
  {pages} {155417} (\bibinfo {year} {2016})}\BibitemShut {NoStop}%
\bibitem [{\citenamefont {Clerk}\ \emph {et~al.}(2010)\citenamefont {Clerk},
  \citenamefont {Devoret}, \citenamefont {Girvin}, \citenamefont {Marquardt},\
  and\ \citenamefont {Schoelkopf}}]{Clerk_2010}%
  \BibitemOpen
  \bibfield  {author} {\bibinfo {author} {\bibfnamefont {A.~A.}\ \bibnamefont
  {Clerk}}, \bibinfo {author} {\bibfnamefont {M.~H.}\ \bibnamefont {Devoret}},
  \bibinfo {author} {\bibfnamefont {S.~M.}\ \bibnamefont {Girvin}}, \bibinfo
  {author} {\bibfnamefont {F.}~\bibnamefont {Marquardt}}, \ and\ \bibinfo
  {author} {\bibfnamefont {R.~J.}\ \bibnamefont {Schoelkopf}},\ }\bibfield
  {title} {\enquote {\bibinfo {title} {Introduction to quantum noise,
  measurement, and amplification},}\ }\href@noop {} {\bibfield  {journal}
  {\bibinfo  {journal} {Rev.\ Mod.\ Phys.}\ }\textbf {\bibinfo {volume} {82}},\
  \bibinfo {pages} {1155} (\bibinfo {year} {2010})}\BibitemShut {NoStop}%
\bibitem [{\citenamefont {Liu}\ \emph {et~al.}(2015{\natexlab{a}})\citenamefont
  {Liu}, \citenamefont {Cheng},\ and\ \citenamefont {Lutchyn}}]{Liu_2015}%
  \BibitemOpen
  \bibfield  {author} {\bibinfo {author} {\bibfnamefont {D.~E.}\ \bibnamefont
  {Liu}}, \bibinfo {author} {\bibfnamefont {M.}~\bibnamefont {Cheng}}, \ and\
  \bibinfo {author} {\bibfnamefont {R.~M.}\ \bibnamefont {Lutchyn}},\
  }\bibfield  {title} {\enquote {\bibinfo {title} {Probing {M}ajorana physics
  in quantum-dot shot-noise experiments},}\ }\href@noop {} {\bibfield
  {journal} {\bibinfo  {journal} {Phys.\ Rev.\ B}\ }\textbf {\bibinfo {volume}
  {91}},\ \bibinfo {pages} {081405(R)} (\bibinfo {year}
  {2015}{\natexlab{a}})}\BibitemShut {NoStop}%
\bibitem [{\citenamefont {Liu}\ \emph {et~al.}(2015{\natexlab{b}})\citenamefont
  {Liu}, \citenamefont {Levchenko},\ and\ \citenamefont {Lutchyn}}]{Liu_2015a}%
  \BibitemOpen
  \bibfield  {author} {\bibinfo {author} {\bibfnamefont {D.~E.}\ \bibnamefont
  {Liu}}, \bibinfo {author} {\bibfnamefont {A.}~\bibnamefont {Levchenko}}, \
  and\ \bibinfo {author} {\bibfnamefont {R.~M.}\ \bibnamefont {Lutchyn}},\
  }\bibfield  {title} {\enquote {\bibinfo {title} {Majorana zero modes choose
  {E}uler numbers as revealed by full counting statistics},}\ }\href@noop {}
  {\bibfield  {journal} {\bibinfo  {journal} {Phys.\ Rev.\ B}\ }\textbf
  {\bibinfo {volume} {92}},\ \bibinfo {pages} {205422} (\bibinfo {year}
  {2015}{\natexlab{b}})}\BibitemShut {NoStop}%
\bibitem [{\citenamefont {Beenakker}(2015)}]{Beenakker_2015}%
  \BibitemOpen
  \bibfield  {author} {\bibinfo {author} {\bibfnamefont {C.~W.~J.}\
  \bibnamefont {Beenakker}},\ }\bibfield  {title} {\enquote {\bibinfo {title}
  {Random-matrix theory of {M}ajorana fermions and topological
  superconductors},}\ }\href@noop {} {\bibfield  {journal} {\bibinfo  {journal}
  {Rev.\ Mod.\ Phys.}\ }\textbf {\bibinfo {volume} {87}},\ \bibinfo {pages}
  {1037} (\bibinfo {year} {2015})}\BibitemShut {NoStop}%
\bibitem [{\citenamefont {Haim}\ \emph {et~al.}(2015)\citenamefont {Haim},
  \citenamefont {Berg}, \citenamefont {von Oppen},\ and\ \citenamefont
  {Oreg}}]{Haim_2015}%
  \BibitemOpen
  \bibfield  {author} {\bibinfo {author} {\bibfnamefont {A.}~\bibnamefont
  {Haim}}, \bibinfo {author} {\bibfnamefont {E.}~\bibnamefont {Berg}}, \bibinfo
  {author} {\bibfnamefont {F.}~\bibnamefont {von Oppen}}, \ and\ \bibinfo
  {author} {\bibfnamefont {Y.}~\bibnamefont {Oreg}},\ }\bibfield  {title}
  {\enquote {\bibinfo {title} {Current correlations in a {M}ajorana beam
  splitter},}\ }\href@noop {} {\bibfield  {journal} {\bibinfo  {journal}
  {Phys.\ Rev.\ B}\ }\textbf {\bibinfo {volume} {92}},\ \bibinfo {pages}
  {245112} (\bibinfo {year} {2015})}\BibitemShut {NoStop}%
\bibitem [{\citenamefont {Smirnov}(2017)}]{Smirnov_2017}%
  \BibitemOpen
  \bibfield  {author} {\bibinfo {author} {\bibfnamefont {S.}~\bibnamefont
  {Smirnov}},\ }\bibfield  {title} {\enquote {\bibinfo {title} {Non-equilibrium
  {M}ajorana fluctuations},}\ }\href@noop {} {\bibfield  {journal} {\bibinfo
  {journal} {New J. Phys.}\ }\textbf {\bibinfo {volume} {19}},\ \bibinfo
  {pages} {063020} (\bibinfo {year} {2017})}\BibitemShut {NoStop}%
\bibitem [{\citenamefont {Smirnov}(2018)}]{Smirnov_2018}%
  \BibitemOpen
  \bibfield  {author} {\bibinfo {author} {\bibfnamefont {S.}~\bibnamefont
  {Smirnov}},\ }\bibfield  {title} {\enquote {\bibinfo {title} {Universal
  {M}ajorana thermoelectric noise},}\ }\href@noop {} {\bibfield  {journal}
  {\bibinfo  {journal} {Phys.\ Rev.\ B}\ }\textbf {\bibinfo {volume} {97}},\
  \bibinfo {pages} {165434} (\bibinfo {year} {2018})}\BibitemShut {NoStop}%
\bibitem [{\citenamefont {Valentini}\ \emph {et~al.}(2016)\citenamefont
  {Valentini}, \citenamefont {Governale}, \citenamefont {Fazio},\ and\
  \citenamefont {Taddei}}]{Valentini_2016}%
  \BibitemOpen
  \bibfield  {author} {\bibinfo {author} {\bibfnamefont {S.}~\bibnamefont
  {Valentini}}, \bibinfo {author} {\bibfnamefont {M.}~\bibnamefont
  {Governale}}, \bibinfo {author} {\bibfnamefont {R.}~\bibnamefont {Fazio}}, \
  and\ \bibinfo {author} {\bibfnamefont {F.}~\bibnamefont {Taddei}},\
  }\bibfield  {title} {\enquote {\bibinfo {title} {Finite-frequency noise in a
  topological superconducting wire},}\ }\href@noop {} {\bibfield  {journal}
  {\bibinfo  {journal} {Physica E}\ }\textbf {\bibinfo {volume} {75}},\
  \bibinfo {pages} {15} (\bibinfo {year} {2016})}\BibitemShut {NoStop}%
\bibitem [{\citenamefont {Bathellier}\ \emph {et~al.}(2019)\citenamefont
  {Bathellier}, \citenamefont {Raymond}, \citenamefont {Jonckheere},
  \citenamefont {Rech}, \citenamefont {Zazunov},\ and\ \citenamefont
  {Martin}}]{Bathellier_2019}%
  \BibitemOpen
  \bibfield  {author} {\bibinfo {author} {\bibfnamefont {D.}~\bibnamefont
  {Bathellier}}, \bibinfo {author} {\bibfnamefont {L.}~\bibnamefont {Raymond}},
  \bibinfo {author} {\bibfnamefont {T.}~\bibnamefont {Jonckheere}}, \bibinfo
  {author} {\bibfnamefont {J.}~\bibnamefont {Rech}}, \bibinfo {author}
  {\bibfnamefont {A.}~\bibnamefont {Zazunov}}, \ and\ \bibinfo {author}
  {\bibfnamefont {T.}~\bibnamefont {Martin}},\ }\bibfield  {title} {\enquote
  {\bibinfo {title} {Finite frequency noise in a normal metal - topological
  superconductor junction},}\ }\href@noop {} {\bibfield  {journal} {\bibinfo
  {journal} {Phys.\ Rev.\ B}\ }\textbf {\bibinfo {volume} {99}},\ \bibinfo
  {pages} {104502} (\bibinfo {year} {2019})}\BibitemShut {NoStop}%
\bibitem [{\citenamefont {Altland}\ and\ \citenamefont
  {Simons}(2010)}]{Altland_2010}%
  \BibitemOpen
  \bibfield  {author} {\bibinfo {author} {\bibfnamefont {A.}~\bibnamefont
  {Altland}}\ and\ \bibinfo {author} {\bibfnamefont {B.}~\bibnamefont
  {Simons}},\ }\href@noop {} {\emph {\bibinfo {title} {Condensed Matter Field
  Theory}}},\ \bibinfo {edition} {2nd}\ ed.\ (\bibinfo  {publisher} {Cambridge
  University Press, Cambridge},\ \bibinfo {year} {2010})\BibitemShut {NoStop}%
\bibitem [{\citenamefont {Averin}(2000)}]{Averin_2000}%
  \BibitemOpen
  \bibfield  {author} {\bibinfo {author} {\bibfnamefont {D.~V.}\ \bibnamefont
  {Averin}},\ }\bibfield  {title} {\enquote {\bibinfo {title} {Quantum
  computing and quantum measurement with mesoscopic {J}osephson junctions},}\
  }\href@noop {} {\bibfield  {journal} {\bibinfo  {journal} {Fortschr. Phys.}\
  }\textbf {\bibinfo {volume} {48}},\ \bibinfo {pages} {1055} (\bibinfo {year}
  {2000})}\BibitemShut {NoStop}%
\bibitem [{\citenamefont {Clerk}\ and\ \citenamefont
  {Stone}(2004)}]{Clerk_2004}%
  \BibitemOpen
  \bibfield  {author} {\bibinfo {author} {\bibfnamefont {A.~A.}\ \bibnamefont
  {Clerk}}\ and\ \bibinfo {author} {\bibfnamefont {A.~D.}\ \bibnamefont
  {Stone}},\ }\bibfield  {title} {\enquote {\bibinfo {title} {Noise and
  measurement efficiency of a partially coherent mesoscopic detector},}\
  }\href@noop {} {\bibfield  {journal} {\bibinfo  {journal} {Phys.\ Rev.\ B}\
  }\textbf {\bibinfo {volume} {69}},\ \bibinfo {pages} {245303} (\bibinfo
  {year} {2004})}\BibitemShut {NoStop}%
\bibitem [{\citenamefont {Mozyrsky}\ \emph {et~al.}(2004)\citenamefont
  {Mozyrsky}, \citenamefont {Martin},\ and\ \citenamefont
  {Hastings}}]{Mozyrsky_2004}%
  \BibitemOpen
  \bibfield  {author} {\bibinfo {author} {\bibfnamefont {D.}~\bibnamefont
  {Mozyrsky}}, \bibinfo {author} {\bibfnamefont {I.}~\bibnamefont {Martin}}, \
  and\ \bibinfo {author} {\bibfnamefont {M.~B.}\ \bibnamefont {Hastings}},\
  }\bibfield  {title} {\enquote {\bibinfo {title} {Quantum-limited sensitivity
  of single-electron-transistor-based displacement detectors},}\ }\href@noop {}
  {\bibfield  {journal} {\bibinfo  {journal} {Phys.\ Rev.\ Lett.}\ }\textbf
  {\bibinfo {volume} {92}},\ \bibinfo {pages} {018303} (\bibinfo {year}
  {2004})}\BibitemShut {NoStop}%
\end{thebibliography}
\end{document}